\documentclass[twocolumn]{article}
\usepackage[noadjust]{cite}
\usepackage{amsmath,amssymb,amsfonts}
\usepackage{algorithmic}
\usepackage{color}
\usepackage{graphicx}
\usepackage{textcomp}
\usepackage[utf8]{inputenc} 
\usepackage[T1]{fontenc}    
\usepackage{booktabs}       
\usepackage{amsfonts}       
\usepackage{microtype}      
\usepackage{lipsum}
\usepackage{comment}
\usepackage{amsmath}
\usepackage{amssymb}
\usepackage{graphicx}
\usepackage{mathrsfs}
\usepackage{amsthm}
\usepackage{dsfont}
\usepackage{tikz}
\newtheorem{lemma}{Lemma}
\newtheorem{remark}{Remark}
\newtheorem{theorem}{Theorem}
\newtheorem{proposition}{Proposition}
\newtheorem{assumption}{Assumption}
\newtheorem{definition}{Definition}

\definecolor{MyGreen}{RGB}{50,140,80}

\newcommand{\overbar}[1]{\mkern 1.5mu\overline{\mkern-1.5mu#1\mkern-1.5mu}\mkern 1.5mu}

\def\RR{\mathbb{R}}
\def\Tin{T_{\rm{in}}}
\def\bTin{\overbar{T}_{\rm{in}}}

\def\xin{x_{\rm{in}}}
\def\bxin{\overline{x}_{\rm{in}}}

\def\bTi+1{\overbar{T}_{\rm{i+1}}}
\def\Ti-1{T_{\rm{i-1}}}
\DeclareMathOperator{\sat}{sat}

\DeclareMathOperator{\rank}{rank}
\DeclareMathOperator{\sgn}{sgn}

\def\cR{\mathcal{R}}
\def\cU{\mathcal{U}}
\def\cV{\mathcal{V}}
\def\xss{{x_{ss}}}
\def\uss{{u_{ss}}}
\def\zss{{z_{ss}}}

\title{Integral action for bilinear systems with application to counter current heat exchanger}
\author{Francesco Ripa,   Daniele Astolfi, Boussad Hamroun  and 
Diego Regruto
	\thanks{This work has been partially funded by  the ANR via the project ALLIGATOR (ANR-22-CE48-0009-01). \\
    F. Ripa and D. Regruto are with the Dipartimento di Automatica e Informatica, Politecnico di Torino, Corso Duca degli Abruzzi 24, 10129 Torino, Italy (e-mail: \texttt{(name.surname)@polito.it)}. \\
    B. Hamroun and D. Astolfi  are with the Univ. Lyon, Universit\'e Claude Bernard Lyon~1, CNRS, LAGEPP UMR 5007, Villeurbanne F-69100, France  (e-mail: \texttt{(name.surname)@univ-lyon1.fr)} }}

\date{}

\begin{document}

\maketitle

\begin{abstract}
In this study, we propose a robust control strategy for a counter-current heat exchanger. The primary objective is to regulate the outlet temperature of one fluid stream by manipulating the flow rate of the  second counter-current fluid stream. By leveraging the energy balance equations, we develop a structured bilinear system model derived by using a uniform spatial discretization of each stream into a cascade of homogeneous volumes and by considering the heat transfer and convective phenomena within the exchanger. We introduce two control strategies: (i) an output feedback controller incorporating a state observer and (ii) a purely integral control law. The effectiveness of the proposed control strategy is validated through real
experiments on a real heat exchanger. 
\end{abstract}


\section{Introduction}

Heat exchangers (HEXs) are fundamental components in systems where thermal energy exchange between two or more fluid streams is required. They play a pivotal role across a wide range of industrial applications, including chemical processing plants \cite{Gu2015}, district heating and cooling networks \cite{SAKAWA2002677}, thermodynamic machinery \cite{WU201632}, as well as applications in the the food and pharmaceutical industries\cite{duroudier2016heat}. Given the increasing industrial demand for improved thermal efficiency and energy savings, the control and optimization of heat exchangers have become topics of significant and growing interest \cite{fratczak2018practical}.

A model for a HEX can be obtained
in the form of a distributed parameter system 
by writing energy balance equations, that is 
a set of partial differential equations (PDE) 
where the state variables are space and time dependent. Several authors addressed the control  of a  HEX based on a PDE model, see, e.g.,  \cite{MAIDI2009,Cassol2019,HUHTALA2019}.
For output temperature control, finite-dimensional approximations are frequently adopted in the literature, as in \cite{Varga1995,Scholten17,Chandrashekar1982}. These models generally fall into two main categories: (i) those based on thermodynamic principles, potentially nonlinear; and (ii) those adopting linear input-output dynamic representations. As a result, the control of HEX systems has been explored through a variety of approaches depending on the chosen model structure. Among these are partial feedback linearization \cite{Alsopt89,maidi2020pde}, nonlinear dynamic output-feedback controllers for simplified bi-compartmental models \cite{ZAVALARIO2009}, and model predictive control (MPC) for nonlinear models \cite{SRIDHAR2016}. PID controllers are also commonly employed in practical applications \cite{DIAZMENDEZ2014}.
  
  In \cite{zitte2020robust}, the authors propose a control strategy for a counter-current heat exchanger (HEX) based on a finite-dimensional model. The HEX is represented as a cascade of homogeneous compartments, and the dynamic model is derived by formulating the energy balance equations for each compartment. These equations account for convective heat transfer, heat exchange between the hot and cold fluid streams, and assume a uniform mass flow rate for both fluids.
  The control law is designed using the forwarding approach, as introduced in \cite{Praly2001, Astolfi2017}, \color{black} and assumes the availability of full state measurements. This requirement renders the strategy implementationally challenging, as the number of necessary sensors increases linearly with the number of compartments used to model the HEX.
  \color{black}

The aim of this work is twofold: to address a concrete control problem of practical relevance, and to place our contribution within the broader theoretical framework of bilinear system control. Specifically, we focus on
\color{black} the problem of output regulation for bilinear systems subject to input saturation, in the presence of 
constant references and disturbances. \color{black} In line with previous works \cite{longchamp2003controller, kawano2023stabilization, monfared2023stabilization, andrieu2012global, chen1998exponential, gutman2003stabilizing}, we assume that the system is open loop stable. 
This condition  is satisfied by many real world applications such as heat exchangers \cite{zitte2020robust} and power flow converters \cite{simon2023robust}. To address this problem, we extend the system by incorporating an integral action and   propose 
\color{black}
two distinct
output-feedback control strategies to stabilize the resulting extended system.

The first strategy relies on a classical separation principle approach, combining a state-feedback controller with a state observer. The state-feedback design is a direct application of the forwarding technique presented in \cite{Astolfi2017}. However, unlike \cite{Astolfi2017}, our method explicitly addresses the bilinear nature of the system and prioritizes practical tunability for engineering applications. Furthermore, unlike \cite{monfared2023stabilization}, our strategy does not rely on the passivity properties of the plant, offering a more general design framework aligned with \cite{kawano2023stabilization}. For the estimation layer, we design a Luenberger observer using the LMI-based approach from \cite{zemouche2013lmi}. Crucially, while previous works such as \cite{sacchelli2020dynamic} successfully employed dynamic observers for output stabilization, their design relied on ``slow'' observers, which are unsuitable in the presence of integral action. In our setting, the dynamics introduced by the integrator necessitate a sufficiently fast estimation to maintain stability. We explicitly address this requirement and rigorously demonstrate the stability of the closed-loop system by constructing a composite Lyapunov function.

Finally, we propose a second strategy based on a pure output-feedback controller combined directly with integral action. While this approach requires simpler implementation, it rests on more restrictive assumptions than the observer-based strategy. We analyze the properties of this controller leveraging tools inspired by singular perturbation methods, as developed in \cite{simpson2020analysis,lorenzetti2022saturating}. This analysis highlights the trade-off between the structural simplicity of the control law and the theoretical conditions required to guarantee its validity.
\color{black}

\color{black}
In the second part of this work, we experimentally validate the proposed control strategies on a physical heat exchanger. Specifically, we benchmark the performance of the observer-based output-feedback strategy against a standard Proportional-Integral (PI) controller. While the provided theoretical analysis  focuses on the stability properties of pure integral action, the experimental comparison employs a PI controller to represent the current industrial standard. Comparison with this standard highlights the significant advantages of the proposed controller.
\color{black}
Specifically formulated for  bilinear dynamical system, our approach ensures stability at both  local and global levels-guarantees that generic linear controllers cannot provide. 
Furthermore, the integration of a state observer proves critical in applications where physical sensors are limited. By enabling accurate reconstruction of the system state, the observer facilitates real-time monitoring and fault diagnosis, thereby enhancing the reliability and maintainability of the control architecture in real-world contexts.

In Section \ref{sec:higlight}, the control problem is introduced and the proposed control laws are described. Section \ref{sec:proof} provides  proof of the control strategies presented earlier. In Section~\ref{sec:model}, the bilinear model of the counter-current heat exchanger is formulated. The experimental tests are then presented and discussed in Section~\ref{sec:simu}, followed by concluding remarks and future research directions in Section~\ref{sec:conclusions}.

\color{black}
\subsubsection*{Notation}
We denote with $\RR$ the set of real numbers.
Given $x\in \RR^n$, we denote with $|x|$ its Euclidean norm. Given a matrix $A\in \RR^{n\times m}$, we denote with 
$|A|$ the matrix norm induced by the Euclidian norm.
\color{black}


\section{\color{black}Output Feedback Regulation} 

\label{sec:higlight}
\subsection{Problem statement}

Consider a (single-input single-output) bilinear system with input saturation of the form:
\begin{equation}
\label{eq:bi_linear_system}
\begin{aligned}
\dot{x}&=Ax+ ( Bx+b) \sat(u) + E\\
e  & = C x-r\\
y&= D x,\\
\end{aligned}
\end{equation}
where  {\color{black}$x(t)$}$\in \RR^{n}$ is the state, {\color{black}$u(t)$}  $\in \RR$ is the control input, 
{\color{black}$e(t)$} $\in \RR$,  is an output to be regulated to zero, 
{\color{black}$r(t)$} $\in \RR$ is a constant reference, 
and $A,B,E,b,C, D$ are constant matrices of appropriate dimensions.
The (possibly asymmetric) saturation function $\sat:\RR\to\RR$ is defined as follows
\begin{equation}\label{saturation_input}
 \sat(s)=\left \{ \begin{array}{lll} 	
	 \overline u & {\rm if}& s \ge \overline u,
	\\
	s \quad &{\rm if} &	
s\in \cU:= [\underline u, \overline u],
	 \\
     \underline u & {\rm if}& s \le \underline u\,.
	\end{array} \right.
\end{equation}
Note that we suppose that the signal $e$ can be  measured and made available for feedback design. Moreover, we suppose that  $y{\color{black}(t)}\in \RR^p$ are some other 
measured outputs available for feedback.

Given a constant reference $r \color{black}\in \RR$, 
the regulation problem  $\lim_{t\to\infty}e(t)=0$ 
{\color{black}with input constraint $u\in \cU$
}
is solved if the trajectories of the plant \eqref{eq:bi_linear_system}
converges to a  steady-state solution
$(\xss,\uss)\color{black}\in \RR^n\times \cU$
satisfying
\begin{equation}\label{eq:regulator_equations}
\begin{aligned}
0 & =  (A+ B \uss)  \xss +b \uss +E
\\
0 & = C\xss -r.
\end{aligned}
\end{equation}
Solving these equations, we obtain the following conditions: 
\begin{equation}\label{eq:input_output_mapping}
 \begin{aligned}
    &\xss = \pi(\uss),\\ 
    &\pi(u_{ss})  := -(A+ B u_{ss})  ^{-1}(b u_{ss} + E ) \\ 
    &r = C\pi(\uss).
 \end{aligned}
\end{equation} \color{black}
We therefore define $\cR=C\pi(\cU)\subset\RR$ as the set of reachable set-points, 
that is, a set of the form 
$\cR = [\underline r, \overline r]$, 
with 
\begin{equation}
\underline r = \min_{u\in \cU}C\pi(u),\quad \overline r = \max_{u\in \cU}C\pi(u).
\end{equation}
In the rest of the article, 
we suppose that the reference $r$ is  chosen in the set $\cR$.
Adopting from now on  the following compact notation
\begin{equation}\label{eq:Fugu_def}
     F_{u}  := A+ Bu \,,
      \qquad
      g_u := B\pi(u)+ b, 
\end{equation}
and given a desired 
reference $r\in \cR$ with
a corresponding\footnote{This selection may be non-unique since the mapping $u\mapsto C\pi(u)$ is surjective by construction, but not necessarily injective.} input steady-state $\uss\in \cU$, 
we define the error system dynamics
\begin{equation}\label{eq:error_dynamics}
\begin{aligned}
    \dot {\tilde x} & = F_{\uss}\tilde x  +(B\tilde x+g_{\uss}) (\sat(u)- \uss)
    \\
    e & = C\tilde x 
\end{aligned}    
\end{equation}
where $\tilde x: = x - \xss $.
\color{black}
We state now the following assumption
\color{black}
for the dynamics in \eqref{eq:error_dynamics}.

\begin{assumption}\label{ass:standing_assumption}The following holds:
\begin{itemize}
       \item[(a)] for any $u\in \cU$, matrix $F_u$ is Hurwitz;
\item[(b)] for any $u\in \cU$,  $CF_u^{-1} g_u\neq 0$.
\end{itemize}
\end{assumption}

A particularity of bilinear systems 
is that in the presence of constant inputs the dynamics becomes fully linear.
Item (a) assumes that such a  linear dynamics 
is Hurwitz. 
{\color{black}{Such an assumption is satisfied in several practically relevant applications, such as heat exchangers \cite{zitte2020robust} and power flow converters \cite{simon2023robust}, and it has been adopted in a number of contributions on bilinear systems, e.g. \cite{longchamp2003controller,kawano2023stabilization,monfared2023stabilization,andrieu2012global,chen1998exponential,gutman2003stabilizing}.}
}

Item (b) of Assumption \eqref{ass:standing_assumption}
requires instead that the DC gain of the transfer function  $ H(s) = C(sI - F_\uss)^{-1} \, g_\uss$
be nonzero.
In turn, this is equivalent to requiring  the transfer function $H(s)$ to have no zeros at the origin.
This assumption is classical in the theory of linear output regulation and is necessary for
the design of an integral action, see, e.g. 
\cite{hepburn1984structurally,Astolfi2017,astolfi2022harmonic}.

We remark that if one select $u= \uss$
 based on previous assumptions and condition \eqref{eq:input_output_mapping}, the point $\xss$ becomes a globally exponentially stable equilibrium for the closed-loop dynamics
$$
\dot x = (A+B\uss)x +b\uss + E.
$$
This can be easily verified by using 
the error dynamics \eqref{eq:error_dynamics}
and noticing that for this dynamics one obtains the linear dynamics
$$
\dot {\tilde x} = F_{\uss}{\tilde x}, 
\qquad e = C\tilde x.
$$
However, even though this simple open loop controller
ensures the regulation objective $\lim_{t\to\infty}e(t) = 0$, such an approach is not robust in the presence of model parameter uncertainties (i.e. small variations of the matrices $A,B,C,b,E$).

As a consequence, our goal is to design an integral-action based strategy in order to robustly regulate the output $e$ to zero, while maintaining all the trajectories bounded.
In particular, {\color{black}we look for a feedback of the form:}
\begin{equation}\label{eq:bi_integral_action}
   \begin{aligned}
    u &= \uss +{\color{black}\phi}
    \\
\dot z &= e.   
   \end{aligned}
\end{equation}
where $\phi$ {\color{black}is a function which can depend only on $z$ and 
the measured outputs $y$.}
It is worth highlighting that if the closed-loop trajectories
of 
\eqref{eq:bi_linear_system},
\eqref{eq:bi_integral_action}, 
reach any equilibrium $(\xss,\zss)$, on such an equilibrium the regulation objective 
$e=0$ is necessarily achieved thanks to the effect of 
the integral action, see, e.g. \cite{Astolfi2017}.
{\color{black}Furthermore, the use of integral action} is also necessary if robustness to small perturbations is sought, see, e.g. \cite{hepburn1984structurally}.

{\color{black}
The objective of this article is to address the problem of designing the function $\phi$ in an output-feedback context, i.e. where only the variables $e,y$ (and consequently $z$) are available for feedback. To this end, we consider two primary feedback designs.
First, we present a design based on the classical separation principle \cite{andrieu2009unifying}, wherein a state-feedback law is first developed and then implemented using state estimates provided by an observer. The state-feedback is based on the forwarding approach \cite{Astolfi2017}, while the observer is designed using standard Lipschitz techniques \cite{zemouche2013lmi}. Second, by strengthening Assumption \eqref{ass:standing_assumption}, we consider a simple integral feedback law of the form $\phi(z)$. This is implemented in practice as a PI controller and serves as a baseline for comparison with the proposed observer-based regulator.
}


\subsection{\color{black} Separation principle for integral feedback}
{\color{black}
We seek a dynamical output-feedback law of the form
\[
\dot{\hat x} = \Sigma(\hat x,z,y), 
\qquad 
u = \uss + \phi(\hat x,z,y),
\]
where $\hat x$ denotes an estimate of the state based on the measured output $y$ and on the integral variable $z$. The function $\Sigma$ will be designed according to a Luenberger-like observer structure, while the function $\phi$ will be obtained by following a forwarding-based state-feedback design.

The state-feedback law first derived  is therefore not intended for direct implementation on the physical system, since the full state vector $x$ is not available for measurement. Its role is primarily methodological: it provides a convenient theoretical benchmark and an intermediate design step that will be exploited in the construction of the above output-feedback regulator. In particular, the corresponding stability result characterizes the closed-loop state-feedback dynamics and serves as a building block for the subsequent observer-based output-feedback design.

Consistent with this hierarchical approach,
we construct a ``forwarding-based'' feedback law following the methodology in \cite{Astolfi2017}.
Specifically, for any} given  $\uss \in \cU$,
let us introduce  matrices $P,M$ as the
solutions to 
\begin{equation}\label{eq:def_PM}
    F_{\uss}^\top P+PF_{\uss}=-2\Upsilon,
    \qquad M=CF_{\uss}^{-1},
\end{equation}
for some positive definite matrix $\Upsilon\succ 0$.
Note that the equations \eqref{eq:def_PM} always admits a solution since $F_\uss$ is Hurwitz by item (a) of Assumption~\ref{ass:standing_assumption}. 
Next, 
consider the following feedback law
\begin{equation}\label{eq:forwarding}
    \begin{aligned}
   & \phi(x,z) =  -\Big(B  (x-\xss) +g_{\uss}\Big)^\top \times
\\
&\qquad \qquad \quad
   \Big[ k_p (x-\xss) ^\top P- k_i\big(z -M (x-\xss) \big) M) \Big]^\top 
\end{aligned}
\end{equation}
with $k_p, k_i$
being positive gains to be tuned.
\color{black} We state the following proposition serving as an intermediate result for the main result of this section. \color{black}

\begin{proposition}\label{thm:forwarding}
 Suppose Assumption~\ref{ass:standing_assumption} holds.
 Given $(r,\uss)\in R\times \cU$
 satisfying
 \eqref{eq:regulator_equations}, 
  the equilibrium $(\xss,0)$, 
  is globally asymptotically stable and locally exponentially stable
for the closed-loop dynamics
\eqref{eq:bi_linear_system}, 
\eqref{eq:bi_integral_action},
\eqref{eq:forwarding},  
for any $k_i>0$ and $k_p>0$.
\end{proposition}

\begin{proof}
    See Section~\ref{sec:forwarding}.
\end{proof}

\color{black}
We stress that the proposed feedback law 
 cannot be implemented unless the full state is measurable; that is, it requires the condition $D = I$
and $y = x$.
 In case $D$ has rank $p<n$
one can resort to 
a state-observer in order to 
estimate $x$ online. 
\color{black}
To this end, we consider the following additional assumption.

\begin{assumption}\label{ass:observability}
  \color{black}There exists $Q= Q^\top \succ0$, $\nu, \epsilon>0$
and $Y\in \RR^{n\times p}$
satisfying the following LMI condition:
 \begin{equation} \label{eq:observer_condition}
 \begin{pmatrix}
     QA +A^\top Q- YD- D^\top Y^\top +(\nu \mu^2+2\epsilon)I & Q
     \\
     Q & -\nu I
 \end{pmatrix}\preceq 0
   \end{equation}
{\color{black}with $\mu = |B| \max\{|\underline u|, |\overline u|\}$.}
\end{assumption}

\color{black}
Note that a necessary (but not always sufficient) condition for the feasability of the LMI 
\eqref{eq:observer_condition} is the observability of the pair $(A,D)$. 
More details are given in \cite[Theorem 4]{zemouche2013lmi}. \color{black}
Based on the previous assumption, we design a  Luenberger-like observer of the following form
\begin{equation} \label{eq:observer_equation}
    \dot {\hat x} = A\hat x+  (B\hat x + b)\sat (u) + L(y- D\hat x) + E
\end{equation}
where the observer gain $L$  is chosen as {\color{black}$L=Q^{-1}Y$, } 
and the feedback gain $\phi$ in \eqref{eq:bi_integral_action} is now selected as 
\begin{equation}\label{eq:control_theorem3}
    \begin{aligned}
   & \phi(\hat x,z) =  -\Big(B  (\hat x-\xss) +g_{\uss}\Big)^\top \times
\\
&\qquad \qquad \quad
   \Big[ k_p (\hat x-\xss) ^\top P- k_i\big(z -M (\hat x-\xss) \big) M) \Big]^\top .
\end{aligned}
\end{equation}
We have then the following result.

\begin{theorem}\label{thm:separation_principle}
Let Assumptions~\ref{ass:standing_assumption},
\ref{ass:observability}.
Then, given $(r,\uss)\in R\times \cU$
 satisfying
\eqref{eq:regulator_equations}, 
  the equilibrium $(x, z, \hat x)=(\xss,0, \xss)$ 
  is globally asymptotically stable and locally exponentially stable
for the closed-loop dynamics
\eqref{eq:bi_linear_system}, 
\eqref{eq:bi_integral_action},
\eqref{eq:observer_equation}
\eqref{eq:control_theorem3}
for any $k_i>0$, $k_p>0$,
and $L=Q^{-1}Y$.
\end{theorem}

\begin{proof}
    See Section~\ref{sec:separation}.
\end{proof}

{\color{black}
\begin{remark}
The Luenberger observer gain \(L = Q^{-1}Y\) obtained from
\eqref{eq:observer_condition} guarantees robust exponential convergence of
the estimation error. Although the eigenvalue placement of \(A - LD\) is not
explicitly parameterized, the scalar variable \(\nu\) in the LMI directly
affects the convergence rate: increasing \(\nu\), while maintaining
feasibility, yields a faster observer dynamics. Hence, the convergence speed
can be increased by selecting \(L\) through the maximization of \(\nu\) subject to the LMI constraints in
\eqref{eq:observer_condition}.
\end{remark}

}




\subsection{Integral gain feedback}

In this section, we consider 
a simple feedback law using only the solely information of $z$
which is able to stabilize the extended system 
\eqref{eq:bi_linear_system}, 
\eqref{eq:bi_integral_action} that is, 
we look for simple integral gain feedback 
of the form
\begin{equation}\label{eq:bi_feedback_sing_pert}
{\color{black}\phi(z)} = -\sgn(CF_u^{-1} g_u) k_i z,   
\end{equation}
with $k_i>0$ to be chosen small enough.
To show the stability of the interconnection
\eqref{eq:bi_linear_system}, 
\eqref{eq:bi_integral_action}
with 
\eqref{eq:bi_feedback_sing_pert}
the following additional assumption is introduced.

\begin{assumption}\label{ass:sing_pert}The following holds:
   \begin{itemize}
      \item[(a)]
      let $\mu=  |B| \max\{|\underline u|, |\overline u|\}$.
      For    for any  $u\in \cU$
       there exists $P= P^\top \succ 0$ and $\nu,\epsilon>0$ satisfying      the following LMI \begin{equation}\label{eq:stability_condition}
 \begin{pmatrix}
     PF_{u} +F_u^\top P+(\nu \mu^2+2\epsilon)I & P
     \\
     P & -\nu I
 \end{pmatrix}\preceq 0.
   \end{equation}
       \item[(b)] 
       let $\cV=[\underline{u}-\overline u,\overline u-\underline u]$.
       Then, $C(F_{u}+Bv)^{-1}g_{u} \neq 0$ for any $(u,v)\in \cU\times \cV$.
    \end{itemize}    
\end{assumption}

We remark that Assumption~\ref{ass:sing_pert}
implies Assumption~\ref{ass:standing_assumption}.
This can be easily seen because 
the LMI \eqref{eq:stability_condition}  implies the matrix $F_u$ 
being Hurwitz for any $u\in \cU$, 
while item (b)
of Assumption~\ref{ass:sing_pert}
implies 
item (b)
of Assumption~\ref{ass:standing_assumption}
when one takes $v= 0 \in \cV$.
\color{black}
Notably, the robustness against $v \in \mathcal{V}$ is required because the constant gain 
in \eqref{eq:bi_feedback_sing_pert} enters the dynamics 
\eqref{eq:error_dynamics} modulated by the state-dependent term $(B\tilde x+g_{uss})$. This contrasts with the strategy \eqref{eq:forwarding}, where the feedback gain inherently includes the term $(B\tilde x + g_{uss})^\top$, structurally ensuring that the integral action always acts in the correct direction.
\color{black}
Based on the previous assumption, we have the following theorem.

\begin{theorem}\label{thm:singular_pert}
 Suppose Assumption  \ref{ass:sing_pert} holds.
  Given $(r,\uss)\in \cR\times \cU$
 satisfying
 \eqref{eq:regulator_equations}, 
there exists $k_i^*>0$ such that
 the equilibrium $(\xss,0)$, 
  is globally asymptotically stable and locally exponentially stable
for the closed-loop dynamics
\eqref{eq:bi_linear_system}, 
\eqref{eq:bi_integral_action},
     \eqref{eq:bi_feedback_sing_pert}
     for any  $k_i\in (0,k_i^*)$.
\end{theorem}

\begin{proof}
    See Section~\ref{sec:sing_pert}.
\end{proof}

{\color{black}
Given 
$P, \epsilon$ satisfying \eqref{eq:stability_condition},
 let $\underline{p}$ and $\overline{p}$ denote the minimum and maximum eigenvalues of $P$, respectively.
An admissible, though conservative, value for the maximum allowed integral gain $k_i^\ast$ is given by
\[
\begin{aligned}
k_i^\ast & = \frac{\epsilon}{3 |C| \,\overline \pi \,\sqrt{\underline p\,\overline p}},
\\
\overline \pi & = \sup_{v\in V} \left|\big[(F_\uss+Bv)^{-1}B v - I \big](F_\uss+Bv)^{-1} g_\uss\right|.
\end{aligned}
\]
}
It is readily seen that at the prize of stringent assumptions, a simpler controller can be obtained.
Indeed the feedback  \eqref{eq:bi_feedback_sing_pert}
is based on a pure integral feedback 
while the feedback presented in the previous sections, e.g. 
\eqref{eq:forwarding} and \eqref{eq:control_theorem3}
requires a more complicated form.
Nonetheless,
since the parameter gain has to be chosen small enough, 
 a limitation of the proposed method is that it   has possibly poorer convergence properties.

{\color{black}
\subsection{Numerical illustration}

Consider  system \eqref{eq:bi_linear_system} with $C = D = (1, 1)$,
$$
A=\begin{pmatrix}-1&0\\0&-2\end{pmatrix}\!,\
B=\begin{pmatrix}0&0\\0&-1\end{pmatrix}\!,
\
b=\begin{pmatrix}1\\-1\end{pmatrix}\!,
\ 
E=\begin{pmatrix}1\\0\end{pmatrix}\!,
$$
and saturation interval $\mathcal U=[0,1]$.  
We verify the applicability of the proposed output-feedback strategies.

First, we verify Assumptions~\ref{ass:standing_assumption} and
\ref{ass:observability}.
Following the
definitions in
\eqref{eq:input_output_mapping} and \eqref{eq:Fugu_def},
we compute:
\begin{equation*}
\begin{aligned}
& F_u =
\begin{pmatrix}
    -1 & 0 \\ 0 & -2-u
\end{pmatrix}\! , \ 
\pi(u)=\begin{pmatrix}u+1\\-\tfrac{u}{2+u}\end{pmatrix}\! ,\
g_u=\begin{pmatrix}1\\-\tfrac{2}{2+u}\end{pmatrix}\! ,
\\
&
M = CF_u^{-1} = \begin{pmatrix}
    -1 & -\tfrac1{2+u}
\end{pmatrix}, 
\quad
CF_u^{-1}g_u = - 1 + \tfrac{2}{(2+u)^2}.
\end{aligned}
\end{equation*}
It is easy to verify that, for any $u\in \mathcal U$,  $F_u$ is Hurwitz and furthermore 
$CF^{-1}g_u <0$. Thus,
Assumption~\ref{ass:standing_assumption} holds.
Moreover, using the defined values for $B$ and $\cU$, we find that $\mu=1$. It can be also verified that the LMI \eqref{eq:observer_condition} is strictly feasible for $Q=I$, $Y= (1, 1)^\top$, and $\nu=1$ (for sufficiently small $\epsilon > 0$). This confirms that  Assumption~\ref{ass:observability} is satisfied.
Consequently, using the gain $L = (1, 1)^\top$,
the observer \eqref{eq:observer_equation} is guaranteed to convergence,
and the output-feedback integral action strategy developed in Theorem \ref{thm:separation_principle} can be applied.

In contrast,
we  evaluate Assumption~\ref{ass:sing_pert}. Given 
$\cU$, the set $\cV$ is computed as 
 $\cV=[-1,1]$.
Condition  (b) of the assumption requires $C(F_u+Bv)^{-1}g_u\neq 0$
 for all $(u,v)\in \mathcal U\times \mathcal V$. Evaluating this expression yields:
$$
C(F_u+Bv)^{-1}g_0
 = 
 -1 +\dfrac{2}{(2+u)(2+u+v)}.
$$
Specifically, for the choice $u=0 \in \mathcal U$ and $v=-1 \in \mathcal V$, the expression becomes zero, thereby violating Assumption \ref{ass:sing_pert} (b). 
Consequently, for this system, there is no theoretical guarantee that the simple integral feedback law \eqref{eq:bi_feedback_sing_pert} will achieve stabilization.

In conclusion, this simple numerical example demonstrates that Assumption \ref{ass:sing_pert} is more restrictive than Assumption \ref{ass:standing_assumption} and therefore, there is no theoretical guarantee that the pure integral action output feedback strategy developed in \eqref{eq:bi_feedback_sing_pert} would work in this scenario.
\color{black}

\section{Proofs} \label{sec:proof}

In this section we prove {\color{black} the proposition and the main theorems} concerning 
the regulation of system
\eqref{eq:bi_linear_system}, 
\eqref{eq:bi_integral_action}.
To this end, we first introduce the following technical results.

\subsection{Stability results}

We recall in this section the following statement of LaSalle's invariance principle from  \cite{praly2022fonctions}, for Lyapunov functions which are not $C^1$ but only locally Lipschitz.

To this end, we recall the definition of Dini derivative of a function. In particular, given 
a continuous function $\varphi:\RR\mapsto \RR$, 
we define its (upper right hand) Dini derivative 
$$
D^+ \varphi(t) = \limsup_{h\to0+} \dfrac{f(t+h)-f(t)}{h}.
$$
Next,  consider a dynamical system of the form 
\begin{equation}\label{eq:sys_lasalle}
        \dot x = f(x), 
\end{equation}
where $x\in \RR^n$ and
the functions  $f$ is locally Lipschitz.
We  recall the definition of zero-state detectability.   

\begin{definition}[Zero-state detectability]
Consider system \eqref{eq:sys_lasalle} and let 
$h:\RR^n\mapsto\RR^p$, with $1\leq p\leq n$, be a continuous function.
The pair $\color{black}(f,h)$
is said to be zero-state detectable if any solution satisfying 
    $h(x(t))=0$ for all $t\geq0$ converges asymptotically to the origin.
\end{definition}

Finally, we have the following stability result.

\begin{theorem}\label{thm:laSalle}
Consider system \eqref{eq:sys_lasalle}.
Suppose there exists a locally Lipschitz function
{\color{black}$V:\RR^n \to\RR$} , class-${\mathcal K}_\infty$ functions $\underline \alpha, \overline\alpha$, a class-${\mathcal K}$ function $\alpha$ 
and a locally Lipschitz function $h:\RR^n\mapsto\RR^p$, 
with $1\leq p\leq n$
such 
that the    
    following conditions hold:
    \begin{align*}
        &\underline\alpha(|x|)\leq V(x)\leq \overline \alpha(|x|), \quad &&\forall x\in \RR^n,
        \\
        &D^+V \leq -\alpha(|h(x)|), \quad&& \forall x\in \RR^n\setminus \{0\}.
    \end{align*}
    Then, if the pair {\color{black}$(f,h)$}
    is zero-state detectable,  the origin of \eqref{eq:sys_lasalle} is globally asymptotically stable (GAS).
\end{theorem}

\begin{proof}
    See Theorem 2.241 in \cite{praly2022fonctions}
\end{proof}

Finally, we conclude 
The following technical lemma will be used in the proofs.

\begin{lemma}\label{lemma:technical}
  Let $\underline u < \overline u$. 
  Then, 
   $
    s\big(\sat(b-s)-b\big) \leq 0
    $  for any $s\in\RR$ and $b\in [\underline u, \overline u]$.
\end{lemma}

\begin{proof}
Let $S(s) := s\big(\sat(b-s)-b\big) $.
    Consider the following three cases.\\
 Case 1: \( \underline u\leq b-s\leq  \overline{u} \)
which is,      \( b - \overline{u}\leq s\leq  b - \underline{u} \). Hence
    \(
    S(s) = -s^2 \leq 0.
    \)\\
    Case 2:
    \( b-s \leq \underline u \) which is
    \( s \geq b-\underline u \). Therefore, 
    $S(s)= s(\underline u-b)$
    Since $b\geq\underline u$, $S(s)\leq0$.\\
     Case 3: 
     \( b-s \geq \overline u \) which is
    \( s \leq b-\overline u \).
  Hence, 
    $S(s)= s(\overline u-b)$. Since $b\leq \overline u$, we get again $S(s)\leq0$
    completing the proof.
\end{proof}

Finally, 
given a positive definite matrix $P=P^\top \succ 0$, we recall 
the following inequalities:
$$
\dfrac{|s^\top P| }{\sqrt{s^\top P s}} \leq \sqrt{\overline p}, 
\quad
-\dfrac{|s|}{\sqrt{s^\top P s}} \leq -\dfrac1{\sqrt{\underline p}},
\quad
\dfrac{|s|}{\sqrt{s^\top P s}} \leq \dfrac{1}{\sqrt{\underline{p}}},
$$
for any $s\in \RR^n$, with 
$\overline p$, resp. $\underline p$, 
denoting the largest, resp. the smallest, eigenvalue of $P$.

\subsection{Proof of Proposition~\ref{thm:forwarding}}
\label{sec:forwarding}

Using the same change of coordinates 
$\tilde x$ introduced in \eqref{eq:error_dynamics}, 
the closed-loop dynamics
\eqref{eq:bi_linear_system}, 
\eqref{eq:bi_integral_action},
\eqref{eq:forwarding},  reads, 
in the new dynamics, as
\begin{equation}
\begin{aligned}
    \dot {\tilde x} & = F_{\uss}\tilde x  +(B\tilde x+g_{\uss}) v
    \\
    \dot z & = C\tilde x 
    \\
    v  & =  \sat(\uss +\phi(\tilde x,z) )- \uss,
    \\
    \phi (\tilde x,z)& = -  (k_p \tilde x^\top P - k_i (z- M\tilde x) M)(B\tilde x+ g_{\uss})
\end{aligned}    
\end{equation}
Now, consider  the change of coordinates
$$
z \mapsto \tilde z := z- M\tilde x
$$
with $M$ defined as in \eqref{eq:def_PM},
which transform the system into 
$$
\begin{aligned}
    \dot {\tilde x} & =  F_{\uss}\tilde x  +(B\tilde x+g_{\uss}) v
    \\
    \dot {\tilde z} & = - M(B\tilde x+g_{\uss}) v
    \\
    v  & =  \sat(\uss +\tilde\phi(\tilde x,\tilde z) )- \uss,
    \\
    \tilde \phi (\tilde x,\tilde z)& =   -(k_p \tilde x^\top P - k_i \tilde z M)(B\tilde x+ g_{\uss})
\end{aligned}
$$
First, consider $k_p>0$.
In this case, we can consider the Lyapunov function 
\begin{equation}
    \label{eq:Vlyapunov}
V = k_p \tilde x^\top P\tilde x + k_i \tilde z^2.
\end{equation}
Its derivative along solutions satisfies
$$
\begin{aligned}
    \dot V &= - 2k_p \tilde x^\top \Upsilon \tilde x - 2\tilde \phi(\tilde x,\tilde z) \big(\sat (\uss+\tilde\phi(\tilde x,\tilde z)) - \uss\big)
    \\
    &\leq - 2k_p \tilde x^\top \Upsilon \tilde x
\end{aligned}
$$
where in the last inequality we used  Lemma~\ref{lemma:technical}.
By applying La Salle's invariance principle, and the fact that 
$Mg_{\uss} =  CF_\uss^{-1}g_\uss \neq 0$, we can conclude that 
the origin is GAS.
Finally, in order to show the local exponential properties of the closed-loop system, we consider the linearization of the dynamics around the origin, given by
$$
\begin{aligned}
    \dot {\tilde x} & =  F_{\uss}\tilde x
    +  g_\uss v
    \\
    \dot {\tilde z} & = - M g_{\uss}
    v
    \\
    v & = -k_p g_{\uss}^\top P \tilde x+ k_i g_{\uss}^\top M^\top \tilde z
\end{aligned}
$$
Taking again the derivative of 
$V$ defined as in \eqref{eq:Vlyapunov}
 gives
$
\dot V = -\tilde x^\top \Upsilon \tilde x  -2v^2 
$.
In view of item (b) of Assumption~\ref{ass:standing_assumption}, $M g_{\uss}\neq 0$
and therefore
there exists some $\epsilon>0$ 
such that $\dot V\leq -\epsilon (|\tilde x|^2 +|\tilde z|^2)$ showing that the origin is LES.

\subsection{Proof of  Theorem~\ref{thm:separation_principle}}\label{sec:separation}

   Consider the following change of coordinates
\begin{equation}
    \begin{aligned}
        x\mapsto \varepsilon &:= \hat x-x \\
        \hat x\mapsto \tilde x &:= \hat x-\xss
    \end{aligned}
\end{equation}   
The system assumes this form:
\begin{align*}
        \dot {\tilde x} & =F_\uss\tilde x  +(B\tilde x+g_\uss) v -LD\varepsilon
        \\
        \dot z & = C\tilde x - C\varepsilon 
        \\
        \dot {\varepsilon} & = (A+\sat(u)B-LD )\varepsilon
        \\
        v &=  \sat(\uss -\phi(\tilde x,z) )- \uss
         \\
        \phi (\tilde x,z)& =   (k_p \tilde x^\top P - k_i (z-  M\tilde x) M)(B\tilde x+g_\uss)
\end{align*}
Next, we change coordinates as follows
$
z \mapsto \tilde z := z- M\tilde x
$
to obtain
\begin{equation}
\label{eq:complete_system_output_feedback}
\begin{aligned}
   \dot {\tilde x} & =F_\uss\tilde x  +(B\tilde x+g_\uss) v -LD\varepsilon
        \\
    \dot {\tilde z} & = - M (B\tilde x+g_\uss) v + (MLD-C)\varepsilon
    \\
      \dot {\varepsilon} & = (A+\sat(u)B-LD )\varepsilon
    \\
    v &=  \sat(\uss -\tilde \phi(\tilde x,\tilde z) )- \uss
    \\
    \tilde \phi (\tilde x, \tilde z)& =   (k_p \tilde x^\top P - k_i  M\tilde z)(B\tilde x+g_\uss).
\end{aligned}
\end{equation}
 Consider the Lyapunov function 
$U = \varepsilon ^\top Q\varepsilon$
with $Q$ satisfying
\eqref{eq:observer_condition}. Its derivative along solutions gives
\begin{equation*}
            \dot U \leq  \varepsilon^\top (Q(A-LD)+(A-LD)^\top Q)\varepsilon +2\sat(u)\varepsilon^\top QB \varepsilon .
\end{equation*}
Using Young's inequality we have
$$
\begin{aligned}
2\sat(u)\varepsilon^\top QB \varepsilon & \leq \tfrac{1}{\nu}\varepsilon^\top Q Q\varepsilon + \nu\sat(u)^2\varepsilon^\top B^\top B\varepsilon
\\
&\leq \tfrac{1}{\nu}\varepsilon^\top Q Q\varepsilon +
\nu \mu^2 \varepsilon^\top I \varepsilon.
\end{aligned}
$$
Applying  Schur's complement to \eqref{eq:observer_condition} gives
$$
Q(A-LC)+(A-LC)^\top Q+\tfrac1\nu QQ+\nu \mu^2 I\preceq -2 \epsilon I.
$$
As a consequence, combining the previous inequalities we finally obtain
$\dot U \leq -2\epsilon|\varepsilon|^2$.

Next, consider the Lyapunov function 
\begin{equation}
\label{eq:Wdefinition_2theorem}    
\begin{aligned}
    W & := \sqrt{V(\tilde  x, \tilde z)} + c\sqrt{U(\varepsilon)}
\\ V&= {k_p \tilde x^\top P\tilde x + k_i\tilde z^2} , 
\quad U = {\varepsilon^\top Q\varepsilon}.
\end{aligned}
\end{equation}
We denote with $\overline q$, resp. $\underline q$, the largest, resp. the smallest, eigenvalue of $Q$.
Following similar computations as in the proof of Proposition~\ref{thm:forwarding} the derivative of $W$ defined in 
\eqref{eq:Wdefinition_2theorem}
along solutions to 
\eqref{eq:complete_system_output_feedback}
yields
$$
\begin{aligned}
   D^+ W &\leq  
  \dfrac{ -k_p \tilde x^\top \Upsilon \tilde x - \tilde \phi(\tilde x, \tilde z) \big(\sat (\uss+ \tilde \phi(\tilde x, \tilde z))-\uss\big)}{\sqrt{V(\tilde x, \tilde z)}}
   \\ & + 
   \dfrac{-k_p\tilde x^\top P LD\varepsilon
+k_i\tilde z(MLD-C)\varepsilon}{\sqrt{V(\tilde x, \tilde z)}}
- c \dfrac{\epsilon |\varepsilon|^2}{\sqrt{U(\varepsilon)}}
    \\
    & \leq  -  \dfrac{ k_p x^\top \Upsilon \tilde x }{\sqrt{V(\tilde x, \tilde z)}} 
    - \bigg(\dfrac{c\epsilon}{\sqrt{\underline q}}  -a \bigg)|\varepsilon|
\end{aligned}
$$
with $a>0$ satisfying
$$
\left|\dfrac{k_p \tilde x^\top P LD
+k_i\tilde z(MLD-C) }{\sqrt{k_p \tilde x^\top P\tilde x + k_i\tilde z^2}}\right|\leq a, 
\quad \forall (\tilde x,\tilde z)\neq 0.
$$
By letting $c> a {\sqrt {\underline q}}/\epsilon$, we obtain
$D^+W \leq - \underline\epsilon(|\tilde x|+|\varepsilon|)$ for any $(\tilde x, \tilde z, \varepsilon)$ for some $\underline\epsilon>0$.  
Invoking Theorem~\ref{thm:laSalle}
we conclude that the origin of \eqref{eq:complete_system_output_feedback} is GAS.
The local analysis follows a similar approach to the state-feedback case. Consequently, the detailed computations are not presented.

\subsection{Proof of Theorem~\ref{thm:singular_pert}}
\label{sec:sing_pert}
Let $\uss$ be fixed in the rest of the proof and  consider the following change of coordinates
\begin{equation*}
    \label{eq:xi-coordinates}
x\mapsto \tilde x := x-x^*
\end{equation*}
which gives: 
\begin{equation}\label{eq:xi_error_coordinates}
\begin{aligned}
\dot {\tilde x}&= F_\uss\tilde x+ (B\tilde x+g_\uss) v
\\    
\dot z  & = C\tilde x
\\
v & =  \sat(\uss - \sgn(CF_\uss^{-1}g_\uss)k_i z)- \uss
\end{aligned}
\end{equation}
By definition, we recall that $v\in \cV=[\underline u - \overline u, \overline u -\underline u] $.
Next, define the continuous mapping 
$\Pi:\RR\to\RR$ defined as
$$
\Pi(v) = - (F_\uss+Bv)^{-1}g_\uss .
$$
With such a definition, we can rewrite the dynamics
\eqref{eq:xi_error_coordinates}
as 
\begin{equation}\label{eq:xi_error_coordinates2}
\begin{aligned}
\dot {\tilde x}&= (F_\uss+ Bv)(\tilde x - \Pi(v)v)
\\    
\dot z  & = C(\tilde x-\Pi(v)v)+C\Pi(v)v
\\
v & =  \sat(\uss -\sgn(C\Pi(0))k_i z)- \uss
\end{aligned}
\end{equation}
Consider $P$ satisfying  \eqref{eq:stability_condition} 
and consider the Lyapunov function 
\begin{equation}
    \label{eq:LyapW}
    \begin{aligned}
    W(\tilde x, z) &=  \sqrt{V(\tilde x, z)}+ \gamma|z|, 
    \\ V(\tilde x, z) &= (\tilde x-\Pi(v)v)^\top P (\tilde x-\Pi(v)v).
    \end{aligned}
\end{equation}
for some $c>0$ to be chosen. In the following, we denote 
 with $\overline p$, resp. $\underline p$, the largest, resp. the smallest, eigenvalue of $P$.
 Since $v = 0$ for $z=0$, it can be verified that $W(0,0)= 0$ and moreover
$$
\underline w (|\tilde x| + | z| )
\leq W(\tilde x, z)\leq \overline w (|\tilde x| + | z| ).
$$
for some  $\overline w > \underline w>0$.
Next, we compute some inequality that we will use in order to compute the derivative of $W$ along solutions to 
\eqref{eq:xi_error_coordinates2}.

First, consider the function $v\mapsto \Pi(v)$. It  is continuous and differentiable. 
By recalling that given a differentiable matrix $\Phi(t)$ one has 
$$
\dfrac{d}{dt} (\Phi(t)^{-1}) = -\Phi(t)^{-1} \dfrac{d \Phi}{dt}(t)\Phi(t)^{-1}
$$
we obtain
$$
\begin{aligned}
\dfrac{d}{dt}\left(\Pi(v) v\right) &=  (\dot\Pi(v)v + \Pi(v) )\dot v
 \\
 & =\big[(F_\uss+Bv)^{-1}B v - I \big](F_\uss+Bv)^{-1} g_\uss\dot v.
\end{aligned}
$$
Furthermore, recalling the definition of $v$ one has
$$
\dfrac{d}{dt}v = \left\{\begin{array}{ll}
   0 
    &  \textnormal{if} 
\; z \geq \dfrac{\uss-\underline u}{k_i\sgn(C\Pi(0))},
    \\
   0 
    &  \textnormal{if} 
\; z \leq \dfrac{\uss-\overline u}{k_i\sgn(C\Pi(0))},
    \\
- k_i\sgn(C\Pi(0)) \dot z & \textnormal{otherwise}.
\end{array}\right.
$$
As a consequence, we have  
$$
\begin{aligned}
   D^+|v| &\leq k_i \, D^+|\dot z| 
   \\
   &\leq  k_i|C(\tilde x- \Pi(v)v)| + k_i|C \Pi(v)v| .
\end{aligned}
$$
Combining together all the previous bounds, one obtains
\begin{equation}
\label{eq:derivPiv}
D^+|\Pi(v)v|\leq k_i \overline \pi c_0|(\tilde x- \Pi(v))v| + k_i \overline \pi|C \Pi(v)v| 
\end{equation}
with  $c_0= |C|$ and 
$$
\overline \pi = \sup_{v\in V} \left|\big[(F_\uss+Bv)^{-1}B v - I \big](F_\uss+Bv)^{-1} g_\uss\right|.
$$
Next, we compute
$$
C\Pi(v)v z = 
C\Pi(v)z\big(\sat(\uss -k_i\sgn(C\Pi(0))z)-\uss\big).
$$
Note that since $\uss\in \cU = [\underline u, \overline u]$ with $\overline u \geq \underline u$, we have $v\in \cV = [\underline u - \overline u, \overline u - \underline u]$. As a consequence, 
since $C\Pi(v)\neq 0$ for all $v\in \cV$,
in view of item (b) of 
Assumption~\ref{ass:sing_pert},
the sign of $C\Pi(v)$ must be constant for all $v\in \cV$. Hence, since $0\in \cV$, we obtain
$C\Pi(v) \sgn(C\Pi(0))>0$.
Using Lemma~\ref{lemma:technical}, we 
therefore obtain 
\begin{equation}
    \label{eq:derivCPiv}
C\Pi(v)v z<0 \qquad\forall v\neq0.
\end{equation}
Then, 
consider the following inequality.
Applying  Schur's complement
and 
 item (a)
of Assumption~\ref{ass:sing_pert}
we have
\begin{align}
 \notag
&s^\top (P (F_\uss+Bv)  +(F_\uss+Bv)^\top P)s
\\  \notag
&\qquad \leq s^\top (PF_\uss+F_\uss^\top P)s
+ \tfrac1\nu s^\top PP s+ \nu s^\top B^\top Bs 
\\
&\qquad \leq -2\epsilon |s|^2 \quad \forall (x,v)\in \RR^n\times V.    \label{eq:ineqRobustStability}
\end{align}
Finally, we can compute the derivative 
of $W$ defined as 
\eqref{eq:LyapW}.
Using inequalities
\eqref{eq:derivPiv},
\eqref{eq:derivCPiv} and 
\eqref{eq:ineqRobustStability},
we obtain
$$
\begin{aligned}
D^+ W &\leq  
\dfrac{(\tilde x- \Pi(v)v)^\top P (F_\uss+ Bv)(\tilde x - \Pi(v)v)}{ \sqrt{V(\tilde x,z)}} 
\\
&+  \dfrac{(\tilde x- \Pi(v)v)^\top P  }{ \sqrt{V(\tilde x,z)}}  D^+ |\Pi(v)v|
+\gamma\,  D^+|z|
\\
&\leq -\dfrac{\epsilon}{\sqrt{\underline p}} \, |\tilde x- \Pi(v)v|
+k_i c_0 \overline \pi\sqrt{\overline p} \, |\tilde x- \Pi(v)v|  \\
&+ k_i \overline \pi \sqrt{\overline p}\, |C\Pi(v)v|
+ \gamma\, c_0 |\tilde x- \Pi(v)v| - \gamma |C\Pi(v)v|.
\end{aligned}
$$
Finally, by selecting
$\gamma = 2k_i \overline\pi \sqrt{\overline p}$
one obtains
$$
\begin{aligned}
D^+ W &\leq - 
\dfrac{1}{\sqrt{\underline p}}\left(\epsilon-3k_i c_0\overline \pi \sqrt{\overline p \underline p} \right)|\tilde x - \Pi(v)v|
\\
& - k_i \overline \pi \sqrt{\overline p} |C\Pi(v)v|
\end{aligned}
$$
Selecting $k_i^* =\epsilon /(3c_0 \overline \pi \sqrt{\underline p\overline p}))$ one get, for any $k_i\in (0,k_i^*)$ the existence of a $\underline \epsilon>0$ such that 
$$
D^+ W \leq -\underline \epsilon\Big(|\tilde x - \Pi(v)v| +|C\Pi(v)v| \Big).
$$
Invoking Theorem~\ref{thm:laSalle}
we conclude that the origin of \eqref{eq:complete_system_output_feedback} is GAS.
Finally, to verify the local properties around the origin, one can verify that the linearization around the origin of 
\eqref{eq:xi_error_coordinates}
is given 
\begin{equation*}
\begin{aligned}
\dot {\tilde x}&= F_\uss\tilde x-g_\uss \sgn(CF_\uss^{-1}g_\uss)k_iz
\\    
\dot z  & = C\tilde x.
\end{aligned}
\end{equation*}
Let $h = CF_\uss^{-1}g_\uss$.
If $\sgn(h)=1$ consider the change of coordinates
$\tilde x\mapsto\xi := \tilde x +k_i F_\uss^{-1}g_\uss z$, otherwise, 
if $\sgn(h)=-1$, consider
$\tilde x\mapsto\xi := \tilde x -k_i F_\uss^{-1}g_\uss z$. 
We develop the computations only in the first case for brevity.
In the new set of coordinates $\chi = (\xi,z)$
we obtain $\dot \chi = {\mathcal A}(k_i)\chi$
with
\begin{equation*}
{\mathcal A}(\epsilon)  =
\begin{pmatrix}
    F_\uss \xi - \epsilon\, F_\uss^{-1} g_\uss C &
   \epsilon^2 F_\uss^{-1} g_\uss h
    \\
    C & -\epsilon \, h
\end{pmatrix}.
\end{equation*}
The matrix $\mathcal A$ is low-Hurwitz stable according to \cite[Appendix II]{simpson2026steady}.
As a consequence, $\mathcal A$ is Hurwitz for a sufficiently small $k_i$, concluding the proof.


\color{black}

\section{Temperature regulation of the counter-current heat exchanger} \label{sec:model}

\subsection{Modelling}

The transport and exchange of thermal energy within the hot and cold fluid streams circulating in the heat exchanger can be accurately modeled by first-order hyperbolic partial differential equations (PDEs) derived from fundamental conservation physical laws \cite{zobiri2017pde,huhtala2019robust}. Heat exchangers are characterized by two primary manipulated variables: the flow rates and the inlet temperatures of the fluid streams. By acting on these variables, one can effectively regulate the outlet temperatures of the fluids. It is important to note that the fundamental control-theoretic properties of the heat exchanger system depend critically on the choice of manipulated variables.

When the inlet temperatures are manipulated, the heat exchanger behaves as a linear distributed parameter system (DPS), allowing for tractable control-theoretic analysis through the application of semigroup theory \cite{maidi2020pde, huhtala2019robust,zobiri2017pde,grabowski2007stability}. Conversely, manipulation via flow rates leads to a nonlinear (bilinear) DPS model for the heat exchanger, thereby complicating both the control design and the theoretical analysis of its properties \cite{buhler2013topics,mechhoud2025adaptive}.

The PDE system of the heat exchanger with saturated control is:
\begin{equation}
\begin{aligned}
\frac{\partial T}{\partial t}(x,t) + \frac{\mathrm{sat}(q(t))}{\rho c_p} \frac{\partial T}{\partial x}(x,t) &= -\frac{\alpha}{\rho c_p} \big( T(x,t) - \overline{T}(x,t) \big), \\
\frac{\partial \overline{T}}{\partial t}(x,t) - \frac{\overline{q}}{\rho c_p} \frac{\partial \overline{T}}{\partial x}(x,t) &= \frac{\alpha}{\rho c_p} \big( T(x,t) - \overline{T}(x,t) \big),
\end{aligned}
\end{equation}
where \( \alpha = \frac{U A}{V} \) is the distributed heat transfer coefficient, with \( U \) the overall heat transfer coefficient, \( A \) the heat exchange area, and \( V \) the volume.
{\color{black}Here, $q(t)$ denotes the manipulated (time-varying) volumetric flow rate of the fluid associated with the temperature profile $T(x,t)$, while $\overline{q}$ is the (constant) volumetric flow rate of the fluid associated with $\overline{T}(x,t)$; the operator $\mathrm{sat}(\cdot)$ models actuator saturation on the manipulated flow rate.}
The boundary conditions are \( T(0,t) = T_{\mathrm{in}}(t) \) (cold fluid inlet) and \( \overline{T}(L,t) = \overline{T}_{\mathrm{in}}(t) \) (hot fluid inlet, counter-current side), while the initial conditions are \( T(x,0) = T_0(x) \) and \( \overline{T}(x,0) = \overline{T}_0(x) \).
For practical purposes, finite-dimensional approximations are commonly employed in the literature to facilitate control design. These models generally fall into two broad categories. The first consists of thermodynamic phenomenological equations, possibly involving nonlinearities, that aim to capture the physical behavior of the system.
The second is based on a linear input-output dynamic representation. 
Both classes of models have their advantages and limitations, and their selection depends on the specific control objectives and the complexity of the system under consideration.
Our model aligns with the one utilized in  \cite{zitte2020robust} which also relies on thermodynamic phenomenological equations. However, we adopt a novel approach in the model derivation.

We consider a counter-current heat exchanger where single-phase hot and cold fluid streams exchange thermal energy. The pressure is assumed to be constant and uniform along the entire exchanger, with no energy accumulation in the separating wall and no heat exchange with the environment. The convection velocity is spatially uniform and treated as a system input, assumed to reach steady-state condition significantly faster with respect to the slower thermal dynamics. Consequently, the model is derived primarily from energy balance equations.

The system is naturally described as a distributed parameter system, with state variables depending on both space and time. In this work, we adopt a spatial discretization of the exchanger. The hot and cold sides are modeled as cascades of $n$ and $\overline n$
  homogeneous and uniform compartments, respectively, as depicted in Figure~\ref{fig:Echangeur}.

\begin{figure}[htbp]
    \centering
    \begin{tikzpicture}[scale=0.75, every node/.style={font=\Large}]
        \def\bw{1.8} 
        \def\bh{1}   

        \draw (0,0) rectangle (\bw, \bh);
        \draw (0,\bh) rectangle (\bw,2*\bh);
        \node at (0.5*\bw, 0.5*\bh) {1};
        \node at (0.5*\bw, 1.5*\bh) {$\overline{1}$};

        \draw (\bw,0) rectangle (2*\bw, \bh);
        \draw (\bw,\bh) rectangle (2*\bw,2*\bh);
        \node at (1.5*\bw, 0.5*\bh) {2};
        \node at (1.5*\bw, 1.5*\bh) {$\overline{2}$};

        \node at (2.75*\bw, 0.5*\bh) {$\cdots$};
        \node at (2.75*\bw, 1.5*\bh) {$\cdots$};

        \draw (3.5*\bw,0) rectangle (4.5*\bw, \bh);
        \draw (3.5*\bw,\bh) rectangle (4.5*\bw,2*\bh);
        \node at (4*\bw, 0.5*\bh) {$n$};
        \node at (4*\bw, 1.5*\bh) {$\overline{n}$};

        \draw[thick] (0,0) -- (0,2*\bh);
        \draw[thick] (4.5*\bw,0) -- (4.5*\bw,2*\bh);

        \draw[thick,->] (-1.0, 0.5*\bh) -- (0, 0.5*\bh);
        \node at (-1.4, 0.5*\bh) {$T_{in}$};
        \draw[thick,->] (4.5*\bw, 0.5*\bh) -- (4.5*\bw + 1.0, 0.5*\bh);

        \draw[thick,->] (4.5*\bw + 1.0, 1.5*\bh) -- (4.5*\bw, 1.5*\bh);
        \node at (4.5*\bw + 1.4, 1.5*\bh) {$\overline{T}_{in}$};
        \draw[thick,->] (0, 1.5*\bh) -- (-1.0, 1.5*\bh);

    \end{tikzpicture}
    \caption{Counter-current exchanger with inlet and outlet heat flux directions.}
    \label{fig:Echangeur}
\end{figure}
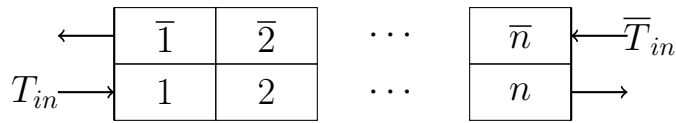
The heat transfer coefficient, denoted by $\lambda$ (J/K/s), is assumed constant. Similarly, the mass density $\rho$ (kg/m$^3$), specific heat capacity $c_p$ (J/kg$\cdot$K), and the compartment volume $V$ (m$^3$) are considered uniform throughout. Unlike earlier models, the present formulation allows for distinct volumes for the hot and cold fluid compartments. The dynamical model is obtained by applying energy balance equations to each compartment, accounting for convective transport and thermal exchange between the streams under the assumption of constant mass flow rates. As previously mentioned, when the mass flow rate is selected as the control input, the resulting system exhibits bilinear dynamics.

These assumptions streamline the modeling procedure while retaining the essential physical properties of the actual system.
The energy balance formulation for each compartment yields a system of differential equations governing the temperature evolution within each unit. For a comprehensive derivation, the reader is referred to~\cite{Zitte2018}.
The thermal dynamics of a heat exchanger, {\color{black}discretized into $n= \overline n$  generic compartments}, are described by the following set of differential equations:
\begin{equation} 
\label{eqbilans}
\left\{
\begin{array}{l}
\dot{T}_1 = \dfrac{\lambda}{\rho V c_p}(\overline{T}_1 - T_1) + \dfrac{q}{\rho V} (T_{\text{in}} - T_1) \\[0.5em]
\dot{T}_i = \dfrac{\lambda}{\rho V c_p}(\overline{T}_i - T_i) + \dfrac{q}{\rho V} (T_{i-1} - T_i), \quad i=2,\dots,n \\[0.5em]
\dot{\overline{T}}_i = -\dfrac{\lambda}{\rho \overline{V} c_p}(\overline{T}_i - T_i) + \dfrac{\overline{q}}{\rho \overline{V}} (\overline{T}_{i+1} - \overline{T}_i), \\
\hfill 
 i=1,\dots,n-1 \\
\dot{\overline{T}}_n = -\dfrac{\lambda}{\rho \overline{V} c_p}(\overline{T}_n - T_n) + \dfrac{\overline{q}}{\rho \overline{V}} (\overline{T}_{\text{in}} - \overline{T}_n)
\end{array}
\right.
\end{equation}
Here, $T_i$ and $\overline{T}_i$
represent the temperatures of the hot and cold fluids, respectively, in the $i$-th compartment. The variables $T_{\text{in}}$ and $\overline{T}_{\text{in}}$ denote the inlet temperatures of the hot and cold streams {\color{black} and are assumed to be different, i.e. $T_{\text{in}} \neq \overline T_{\text{in}} $}, ensuring nontrivial heat transfer. The variables $q$ and $\overline{q}$ are the respective mass flow rates (in kg/s). The parameter $\lambda$ denotes the heat transfer coefficient (J/K/s), while $\rho$ and $c_p$ represent the mass density (kg/m$^3$) and specific heat capacity (J/kg$\cdot$K) of the fluids, assumed identical for both streams. The volumes $V$ and $\overline{V}$ refer to the fluid volumes of a single compartment on the hot and cold side, respectively.

\subsection{Formal verification of the assumptions}

Equation~\eqref{eqbilans} admits a compact matrix representation given by Equation~\eqref{eq:bi_linear_system}. The state vector $x \in \mathbb{R}^{2n}$ is defined as the stacking of the temperatures of the hot and cold fluid compartments, namely $x = (T_1, \dots, T_n, \overline{T}_1, \dots, \overline{T}_n)^\top$.
{\color{black}Without loss of generality, we select the cold side flow rate as the control input \( u= q\in \RR \), while the hot side outlet temperature \( \overline{T}_1\in \RR \) is taken as the controlled output, namely $y = \bar T_1$.} 
\color{black} 
Next, by letting $I_n \in \mathbb{R}^{n \times n}$ be the identity matrix, 
we set
$$
S =
\begin{bmatrix}
-1 & 0 & 0 & \cdots & 0 \\
1 & -1 & 0 & \ddots & \vdots \\
0 & 1 & -1 & \ddots & 0 \\
\vdots & \ddots & \ddots & \ddots & 0 \\
0 & \cdots & 0 & 1 & -1
\end{bmatrix}\in\mathbb{R}^{n\times n}, 
\qquad 
k = \dfrac{\lambda}{\rho c_p},
$$
\[
\overbar A =\begin{bmatrix}
   \overbar{A}_{11} & \overbar{A}_{12}
   \\
   \overbar{A}_{21} & \overbar{A}_{22}
\end{bmatrix}
\quad 
\begin{array}{ll}
\overbar{A}_{11} = -\dfrac{k}{V} I_n, & \quad \overbar A_{12} = \dfrac{k}{V} I_n ,\\
\overbar A_{21} = \dfrac{k}{\overbar{V}} I_n, & \quad \overbar A_{22} = -\dfrac{k}{\overbar{V}} I_n,
\end{array}
\]
\[
\overbar{B} = \dfrac{1}{\rho \overline{V}}
\begin{bmatrix}
0 & 0\\
0 & S^\top
\end{bmatrix}, 
\quad 
b_1 = \dfrac{1}{\rho V}
\begin{bmatrix}
\mathbf{e}_1 \\
0
\end{bmatrix},
\quad
\overline{b}_1 = \dfrac{1}{\rho \overline{V}}
\begin{bmatrix}
0 \\
\mathbf{e}_n
\end{bmatrix},
\]
where $\mathbf{e}_1 = [1\,\, 0\,\, \cdots\,\, 0]^\top$ and $\mathbf{e}_n = [0\,\, \cdots\,\, 0\,\, 1]^\top$ are unit vectors, and $0 \in \mathbb{R}^n$ is the null vector.
It follows that the system matrices $A, B, C, D, E$ and the vector $b$ can be expressed as:
\begin{equation}
\label{eq:hh_matrix_def2} 
A = \overline{A} + \overline{q} \overline{B},
\quad
B = \dfrac{1}{\rho V}
\begin{bmatrix}
S& 0 \\
0& 0
\end{bmatrix},
\quad
b = b_1 T_{\text{in}}, \quad
E = \overline{b}_1 \overline{T}_{\text{in}} \overline{q}.
\end{equation}
The system output, defined as the outlet temperature of the non-manipulated fluid flow ($\overline{T}_{1}$), can be expressed in a compact form as:
\begin{equation}\label{eq:Doutputmap}
y = C x, 
\quad D = C = \begin{bmatrix}
    0^\top   & \mathbf{e}_1^\top  
\end{bmatrix}.
\end{equation}
In our case, the tracking error is computed on the same output, hence $e = y - r = D x - r$, and therefore $C = D$.
\color{black}




At this stage, we verify Assumptions~\ref{ass:standing_assumption} 
aneeded for the application of 
Theorem~\ref{thm:separation_principle}.
 With the mass flow rate \( u(t) = q(t) \), representing the cold fluid as the manipulated control input, the remaining degrees of freedom, namely \( \overline{q} \), \( T_{\text{in}} \), and \( \overline T_{\text{in}} \), are assumed to be fixed at nominal constant values corresponding to steady operating conditions. 
The saturation function 
in \eqref{saturation_input}
is defined with the set $\cU$ satisfying
\begin{equation} \label{eq:setU}
   \cU := [\underline{u}, \overline{u}] \subset \mathbb{R}, \quad \underline u>0 ,
\end{equation}
where \( \underline{u} \) and \( \overline{u} \) denote the minimum and maximum admissible flow rates, respectively.

{\color{black}
We explicitly collect the operating conditions of the heat-exchanger model in the following assumption, which will be used throughout the subsequent analysis.
\begin{assumption}[Operating conditions for the heat exchanger]
\label{ass:operating_condition}
Consider the heat-exchanger model \eqref{eqbilans} operating under admissible conditions satisfying
\[
\lambda > 0, \qquad \overline q > 0, \qquad q = u > 0, \qquad
T_{\mathrm{in}} \neq \overline T_{\mathrm{in}}.
\]
\end{assumption}}
We have now the following result concerning the 
{\color{black}properties} of the matrix \( F_u \).

\begin{lemma}\label{lemma:heat_exch_metzler_hurwitz}
Under Assumption~\ref{ass:operating_condition}, the matrix \( F_{u} = A + B u \)  is Hurwitz {\color{black}and Metzler}, with $A,B$ defined as
    in  \eqref{eq:hh_matrix_def2}. In other words, Assumption~\ref{ass:standing_assumption}(a) holds.
\end{lemma}


{
}

{\color{black}
\begin{proof}
Given any $u\in \cU$, the matrix $F_u$ is defined as
\begin{equation}\label{eq:Fu_proof}
F_{u} =
\begin{bmatrix}
- \Lambda & \Lambda \\
\overline \Lambda  & -\overline \Lambda
\end{bmatrix}
+ \frac{u}{\rho V}
\begin{bmatrix}
S & 0 \\
0 & 0
\end{bmatrix}
+ \frac{\overline{q}}{\rho \overline{V}}
\begin{bmatrix}
0 & 0 \\
0 & S^\top
\end{bmatrix},
\end{equation}
where
\[
\Lambda := \frac{k}{V} I_n,\qquad \overline{\Lambda} := \frac{k}{\overline V} I_n.
\]
Let $\alpha := \frac{k}{V}$ and $\beta := \frac{k}{\overline V}$, and define
\[
P = \mathrm{diag}\!\left(I_n,\frac{\alpha}{\beta}I_n\right)\succ 0,
\qquad
\Upsilon_u := PF_u + F_u^\top P.
\]
Set $L:=S+S^\top$. Straightforward computations yield
\begin{equation}\label{eq:Upsilon_proof}
\Upsilon_u =
\begin{bmatrix}
-2\alpha I_n+ \frac{u}{\rho V}L & 2 \alpha I_n\\
2 \alpha I_n &
-2 \alpha I_n + \frac{\alpha}{\beta}\frac{\overline q}{\rho \overline V} L
\end{bmatrix}.
\end{equation}
The leading principal block of $\Upsilon_u$ satisfies
\begin{equation}\label{eq:lead_principal}
-2\alpha I_n + \frac{u}{\rho V}(S + S^\top) \prec 0 \quad \forall u\in \cU,
\end{equation}
since $\alpha>0$, $S+S^\top=L\prec 0$ and $u\ge 0$.
Applying the Schur complement to~\eqref{eq:Upsilon_proof}, we further obtain
\begin{equation}\label{eq:schur_step}
-2\alpha I_n + \frac{u}{\rho V}L
- 4 \alpha^2 \left(
-2 \alpha I_n + \frac{\alpha}{\beta}\frac{\overline q}{\rho \overline V} L
\right)^{-1} \prec 0.
\end{equation}
Rearranging the expression becomes
\begin{equation}\label{eq:rearranged}
\left[-2\alpha I_n + \frac{u}{\rho V}L\right]
\left[-2 \alpha I_n + \frac{\alpha}{\beta}\frac{\overline q}{\rho \overline V} L\right]
- 4 \alpha^2 I_n \succ 0.
\end{equation}
Expanding~\eqref{eq:rearranged} leads to
\begin{equation}\label{eq:expanded}
4 \alpha^2 I_n
- 2\alpha \left( \frac{u}{\rho V} + \frac{\alpha}{\beta}\frac{\overline q}{\rho \overline V} \right)L
+ \frac{u}{\rho V}\frac{\alpha}{\beta}\frac{\overline q}{\rho \overline V} L^2
- 4 \alpha^2 I_n
\succ 0,
\end{equation}
which simplifies to
\begin{equation}\label{eq:simplified}
-2\alpha \left( \frac{u}{\rho V} + \frac{\alpha}{\beta}\frac{\overline q}{\rho \overline V} \right)L
+ \frac{u}{\rho V}\frac{\alpha}{\beta}\frac{\overline q}{\rho \overline V} L^2
\succ 0.
\end{equation}
Since $L=S+S^\top\prec 0$, the first term in~\eqref{eq:simplified} is positive definite
(because its scalar coefficient is strictly positive for $\overline q>0$ and $u\ge 0$),
and the second term is positive semidefinite, as $L^2\succeq 0$ and all scalar coefficients are nonnegative.
Therefore~\eqref{eq:simplified} holds, which implies $\Upsilon_u\prec 0$ for all $u\in \cU$.
Hence $F_u$ is Hurwitz for any $u\in \cU$.
Moreover, since $\underline u > 0$ by hypothesis, the transport contributions
$(u/\rho V)S$ and $(\overline q/\rho\overline V)S^\top$ have nonnegative off-diagonal entries, the exchange blocks
add only nonnegative cross-couplings, hence $F_u$ is Metzler for all $u\in\cU$.
\end{proof}
}

{\color{black}
\begin{lemma}[Strictly monotone steady-state profile]\label{lem:HX_monotone_ss}
Consider the heat-exchanger model \eqref{eqbilans} under Assumption~\ref{ass:operating_condition}.
Let 
$u\mapsto T_{ss}(u)=(T_{1,ss}(u),\dots,T_{n,ss}(u))\in \RR^n$ 
be the steady-state
mapping 
satisfying equations
\eqref{eqbilans} with $\dot T_i = \dot {\bar T}_i = 0$ for all $i\in \{1, \ldots, n\}$.
Then, $T_{ss}(u)$
is component-wise strictly monotone, namely 
\begin{equation}
T_{\rm in}< T_{1,ss}(u)< \cdots < T_{n,ss}(u).
\end{equation}
\end{lemma}

\begin{proof}
Define the positive constants
\[
a=\frac{\lambda}{\rho V c_p},\qquad
\overline a=\frac{\lambda}{\rho\overline V c_p},\qquad
b=\frac{u}{\rho V},\qquad
\overline b=\frac{\overline q}{\rho\overline V}.
\]
At steady state, \eqref{eqbilans} gives for $i=1,\dots,n$
\begin{equation}\label{eq:ss_hot}
a(\overline T_{i,ss}-T_{i,ss})+b(T_{i-1,ss}-T_{i,ss})=0,
\end{equation}
with $T_{0,ss}=T_{\rm in}$, and
\begin{equation}\label{eq:ss_cold}
-\overline a(\overline T_{i,ss}-T_{i,ss})+\overline b(\overline T_{i+1,ss}-\overline T_{i,ss})=0,
\end{equation}
with $\overline T_{n+1,ss}=\overline T_{\rm in}$.
Let $\Delta T_i:=T_{i-1,ss}-T_{i,ss}$. From \eqref{eq:ss_hot},
\[
\overline T_{i,ss}-T_{i,ss}=-\frac{b}{a}\Delta T_i.
\]
Substituting into \eqref{eq:ss_cold} yields
\[
\overline T_{i+1,ss}-\overline T_{i,ss}=-\kappa\Delta T_i,
\qquad
\kappa:=\frac{\overline a}{\overline b}\frac{b}{a}=\dfrac{u}{\overline q}>0.
\]
On the other hand,
\[
\overline T_{i,ss}=T_{i,ss}-\frac{b}{a}\Delta T_i,
\]
so
\[
\overline T_{i+1,ss}-\overline T_{i,ss}
= -\Delta T_{i+1}-\frac{b}{a}(\Delta T_{i+1}-\Delta T_i).
\]
Equating the two expressions gives the linear recursion
\[
\Delta T_{i+1}=\eta\,\Delta T_i,
\qquad
\eta:=\frac{\frac{b}{a}+\kappa}{1+\frac{b}{a}}>0.
\]
Hence $\Delta T_i=\eta^{\,i-1}\Delta T_1$ for all $i$, so all $\Delta T_i$ share the same sign.
Since $\Delta T_1=T_{\rm in}-T_{1,ss}(u)$, this proves monotonicity of $T_{ss}(u)$, namely 
$\Delta T_i \geq0$
for all $i$. 
Now suppose  $\Delta T_1=0$. As a consequence, 
we obtain $\Delta T_i=0$ for all $i$, implying
$T_{i,ss}=T_{\rm in}$ and, from \eqref{eq:ss_hot}, $\overline T_{i,ss}=T_{\rm in}$ for every $i$.
Substituting into the last cold-stream balance yields $\overline T_{\rm in}=T_{\rm in}$, contradicting the assumption 
$T_{\rm in}\neq \overline T_{\rm in}$ in the statement.
Thus $\Delta T_1\neq 0$, and since $\eta>0$, all $\Delta T_i\neq 0$ with common sign, proving strict monotonicity and concluding the proof.
\end{proof}

\begin{remark}\label{rem:HX_monotone_phys}
Lemma~\ref{lem:HX_monotone_ss} formalizes a behavior that is physically expected in heat exchangers:
in steady state, unidirectional transport coupled with local dissipative heat exchange does not generate oscillatory
spatial profiles; the actuated-stream temperature varies monotonically (cooling or heating depending on the regime).
\end{remark}

}

{\color{black}
\begin{lemma}\label{lem:HX_ass_b_ieee}
Consider the heat-exchanger model \eqref{eqbilans} under Assumption~\ref{ass:operating_condition} and regulated output $y=\overline T_1$.
Then Assumption~\ref{ass:standing_assumption}(b) holds, namely
\[
CF_u^{-1}g_u\neq 0 \qquad \forall u\in(0,\overline u].
\]
\end{lemma}

\begin{proof}
Since $C=\begin{bmatrix}0^\top & e_1^\top\end{bmatrix}$, one has
\[
CF_u^{-1}g_u
=
\sum_{j=1}^{2n} (F_u^{-1})_{n+1,j}\,(g_u)_j ,
\]
with $(g_u)_j$
denoting the $j$-th 
component of the vector  $g_u$ 
which is defined as 
\[
g_u =
\begin{bmatrix}
g_u^{(1)} \\[1mm]
0
\end{bmatrix} \in \mathbb{R}^{2n}, 
\quad  g_u^{(1)} =
\begin{bmatrix}
\dfrac{T_{\rm in} - T_{1,ss}(u)}{\rho V} \\[1mm]
\dfrac{T_{1,ss}(u) - T_{2,ss}(u)}{\rho V} \\[1mm]
\vdots \\[1mm]
\dfrac{T_{n-1,ss}(u) - T_{n,ss}(u)}{\rho V}
\end{bmatrix}.
\]
Note that $g_u$
has nonzero entries only along the actuated temperature chain $T_i$.
Furthermore, by Lemma~\ref{lem:HX_monotone_ss}, the steady-state profile $T_{ss}(u)$ is monotone and nonconstant whenever
$T_{\rm in}\neq \overline T_{\rm in}$ and $u>0$.
Hence all nonzero components of $g_u$ share the same nonzero sign.
Recall that by Lemma~\ref{lemma:heat_exch_metzler_hurwitz} $F_u$ is Metzler and Hurwitz.
As a consequence, 
recalling 
\cite[Theorem B20]{bullo2022contraction}
Therefore $-F_u^{-1}\ge 0$ element-wise, i.e., $F_u^{-1}$ is element-wise non-positive.
It follows that each term $(F_u^{-1})_{n+1,j}(g_u)_j$ in the above sum has the same sign or is zero.
Hence it suffices to show that at least one term is nonzero.
Consider $j=1$, corresponding to $T_1$.
Since $(F_u)_{n+1,1}=k/\overline V>0$ and $F_u$ is Metzler and Hurwitz, one has $(F_u^{-1})_{n+1,1}\neq 0$.
Together with $(g_u)_{1}\neq 0$, this implies that at least one term in the sum is nonzero, concluding 
$CF_u^{-1}g_u\neq 0$
for all $u\in(0,\overline u]$.
\end{proof}

Next, we analyze the domain of the admissible constant reference outputs.

\begin{lemma}
    Consider  the heat-exchanger model 
    \eqref{eqbilans}
    with matrices 
    $A,B,C,E,b$ defined as in \eqref{eq:hh_matrix_def2}, under Assumption~\ref{ass:operating_condition}.
    The mapping 
    $u\mapsto r(u): = C\pi(u)$    as defined in    \eqref{eq:input_output_mapping} is monotone and, for any non-empty set $\cU = [\underline u, \bar u]$, with $0<\underline u<\bar u$, the corresponding set
    of references
    $\cR = C\pi(\cU)$ is non empty.
\end{lemma}

\begin{proof}
    Let $\psi(u):=C x_{ss}(u)=C\pi(u)$. Differentiating the steady-state regulator equation
\[
F_u\pi(u)+bu+E=0
\]
with respect to $u$ gives
\[
\psi'(u)=C\pi'(u)=-CF_u^{-1}\big(B\pi(u)+b\big)=-CF_u^{-1}g_u.
\]
By Lemma~\ref{lem:HX_ass_b_ieee}, $\psi'(u)=-CF_u^{-1}g_u\neq 0$.
As a consequence, 
the steady-state map $\psi(u)=C x_{ss}(u)$ is strictly monotone on $(0,\overline u]$, concluding the proof.
\end{proof}

From the previous lemma, we remark that  the direction of monotonicity depends on the operating regime; for instance, in heating conditions an increase of the
manipulated flow yields an increase of the outlet temperature $y=\overline T_1$.
}

{\color{black}
To conclude, we address Assumption~\ref{ass:observability} for the heat-exchanger model \eqref{eqbilans}.
For the specific set of parameters considered in this work, we have numerically verified the feasibility of the LMI
\eqref{eq:observer_condition}, and thus obtained an observer gain $L=Q^{-1}Y$ satisfying the required condition.
Moreover, the following lemma shows that the pair $(A,D)$ associated with \eqref{eqbilans} is observable (hence detectable)
for the whole class of compartmental heat-exchanger models considered here, thereby validating the necessary structural
condition underlying \eqref{eq:observer_condition}.

\begin{lemma}[Observability of $(A,D)$ for the compartmental heat exchanger]\label{lem:HX_observability}
The pair $(A,D)$ defined in  
\eqref{eq:hh_matrix_def2} and \eqref{eq:Doutputmap} is observable.
\end{lemma}

\begin{proof}
Let $r_k:=DA^k\in\mathbb{R}^{1\times 2n}$ and partition
$r_k=[\,p_k\mid q_k\,]$ according to $x=(T,\overline T)$, with
$p_k,q_k\in\mathbb{R}^{1\times n}$.
From $r_0=D$ we have $p_0=0^\top$ and $q_0=e_1^\top$.
Using $r_{k+1}=r_kA$ and the block structure of $A$, one obtains the recursion
\begin{equation}\label{eq:HX_recursion_D}
\begin{aligned}
p_{k+1} &= -a p_k+\bar a q_k,\\
q_{k+1} &= a p_k-\bar a q_k+\bar b q_k S^\top,
\end{aligned}
\end{equation}
where $a=\frac{k}{V}>0$, $\overline a=\frac{k}{\overline V}>0$, $\overline{} b =\frac{\overline q}{\rho\overline V}>0$, and
$S^\top$ is upper bidiagonal with $(S^\top)_{ii}=-1$, $(S^\top)_{i,i+1}=1$.
We note that right-multiplication by $S^\top$ can shift the support of a row vector by at most one index to the right.
Combining this observation with \eqref{eq:HX_recursion_D} and induction shows that
$q_k$ has zero entries for indices $j\ge k+2$, while $p_k$ has zero entries for indices $j\ge k+1$.

We now track the first nonzero coefficients.
For $i=1,\dots,n$, the coefficient of $\overline T_i$ in $r_{i-1}$ equals $\bar b ^{i-1}$.
The case $i=1$ follows from $r_0=D$.
Assume the claim holds for some $i\in\{1,\dots,n-1\}$.
By the support property above, the $(i+1)$-th entry of $a p_{i-1}-\overline a  q_{i-1}$ is zero, hence
\eqref{eq:HX_recursion_D} yields
\[
(q_i)_{i+1}
=\overline b(q_{i-1}S^\top)_{i+1}
=\overline b(q_{i-1})_i
=\overline b^i.
\]
An analogous argument gives, for $i=1,\dots,n$,
\[
(p_i)_i=\overline a(q_{i-1})_i=\overline a \overline b^{i-1},
\]
because $(p_{i-1})_i=0$ by the support property.

Consider the observability matrix
\[
\mathcal O=\begin{bmatrix}D\\ DA\\ \vdots\\ DA^{2n-1}\end{bmatrix}.
\]
The $n\times n$ submatrix formed by rows $r_0,\dots,r_{n-1}$ and columns
$(\overline T_1,\dots,\overline T_n)$ is lower triangular with diagonal
$(1,\overline b,\dots,\overline b^{n-1})$, hence nonsingular.
Using these rows as pivots, elementary row operations can zero out the columns
$(\overline T_1,\dots,\overline T_n)$ in the remaining rows without changing $\rank(\mathcal O)$.
The resulting $2n\times 2n$ submatrix is block lower triangular, with diagonal blocks having diagonals
$(1,\overline b,\dots,\overline b^{n-1})$ and $(\overline a,\overline a \overline b,\dots,\overline a\overline b^{n-1})$, both nonsingular.
Therefore $\rank(\mathcal O)=2n$ and the pair $(A,D)$ is observable.
\end{proof}

\begin{remark}\label{rem:HX_application}
The previous lemmas verify Assumption~\ref{ass:standing_assumption} for the heat-exchanger model and ensure that the set
of admissible constant references is nonempty. Together with Lemma~\ref{lem:HX_observability} and the numerical
feasibility of the LMI \eqref{eq:observer_condition} for the parameters considered in this work, all hypotheses of
Theorem~\ref{thm:separation_principle} are satisfied. Hence, the outlet temperature $y$ can be regulated to any
admissible
reference $r\in(\underline y,\overline y)\in\cR$ by the output-feedback law \eqref{eq:control_theorem3}. In particular, the
forwarding-based state feedback and the associated observer are obtained by specializing the general design to the
heat-exchanger matrices in \eqref{eq:hh_matrix_def2}, yielding an explicit realization of the proposed dynamical
regulator for the considered plant.
\end{remark}
}


\section{Experimental results} \label{sec:simu}
In this section, we present the experimental results obtained by applying our output feedback controller \eqref{eq:control_theorem3} to a real heat exchanger. The experimental setup includes a PIGNAT heat exchanger.

\begin{figure*}[ht]
    \begin{minipage}[b]{0.35\textwidth}
        \centering
        \includegraphics[width=\textwidth]{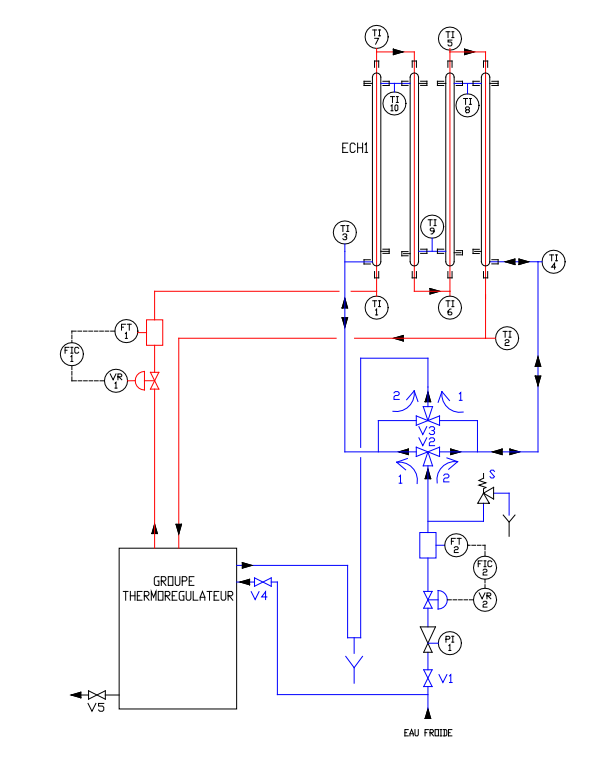}
        \caption{Schematic representation of the PIGNAT Heat Exchanger.}
        \label{fig:schemeHEX}
    \end{minipage}
    \hfill
    \begin{minipage}[b]{0.35\textwidth}
        \centering
        \includegraphics[width=\textwidth]{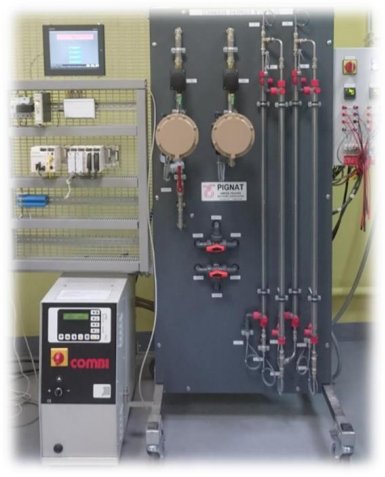}
        \caption{Actual Heat Exchanger.}
        \label{fig:scambiatore_di_calore}
    \end{minipage}
\end{figure*}

Figure~\ref{fig:schemeHEX} shows the schematic representation of the heat exchanger, while figure~\ref{fig:scambiatore_di_calore} depicts the corresponding physical system.
 These images provide a comprehensive overview of the components, including the coaxial pipes housing the two fluids, the reference pumps, and the PLC, which are integrated into the heat exchanger system. Numerical values of the physical parameters are presented in Table 1. For control, we use the cold stream flow rate as the input variable \( u \), and the output temperature of the hot stream, \( \overline{T}_1 \), as the controlled output.

\begin{center}
\begin{table}[h!]
\begin{center}
{\renewcommand{\arraystretch}{1.3}
 \begin{tabular}{|c | c |} 
 \hline
$\lambda = 35\,J/K/s$ & $\rho = 1000\,Kg/m^3$  \\
\hline  
$V = 5.03 \times 10^{-5}\,m^3$ & $\overline{V} = 7.07 \times 10^{-4}\,m^3$ \\
\hline
$c_p = 4186\,J/Kg/K$ (for water) & $\overline{q}=0.02Kg/s$  \\ 
\hline
$\underline{u} = 0\,Kg/s$ & $\overline{u} = 0.05\,Kg/s$ \\
\hline
$\Tin=\xin= 286\,K$ & $\bTin=\bxin = 307\,K$ \\
\hline
\end{tabular}}
\end{center} \vspace{.5em}
\caption{Values of the parameters of the HEX}
\label{tab1}
\end{table}
\end{center}

\subsection{First experiment}
 In this first experiment, we consider 16 compartments, corresponding to $n = 8$ compartments for the hot fluid and $\overline{n} = 8$ compartments for the cold fluid. The control gains were set to $k_p = 0.1 10^-5$ and $k_i =2.6*10^-5$.
The observer gain $L$ was computed by solving the LMI condition in \eqref{eq:observer_condition}, using a standard numerical implementation in \textsc{Matlab}.

The controller’s performance is evaluated by varying the reference temperature. As shown in Figure~\eqref{fig:first_exp}, the reference is initially set to \( 26.5^\circ\mathrm{C} \) for the first 180 seconds. It is then reduced to \( 25^\circ\mathrm{C} \) between 180 and 600 seconds, and subsequently increased to \( 27^\circ\mathrm{C} \) after 600 seconds.
The results demonstrate that the controller exhibits a smooth and stable response, without any sign of actuator saturation. {\color{black}Moreover, at 950 seconds, a \( 0.5^\circ\mathrm{C} \) disturbance is introduced at the output by superimposing a virtual additive perturbation on the measured temperature signal in the Simulink model, so as to emulate a sensor disturbance.} The controller effectively compensates for this disturbance, swiftly restoring the system to the desired temperature.
In Figure~\eqref{fig:first_exp_part2}, the behavior of the observer is presented. The real heat exchanger system is equipped with five sensors, while our discretization consists of 16 compartments. For comparison, we calculated an average value approximately every 3 to 4 blocks and compared it with the measurements from the physical sensors.
Furthermore, the second subfigure of Figure~\eqref{fig:first_exp_part2}, which displays the difference between the true output and the estimated output, confirms that the estimation error is effectively zero.

\begin{figure*}[htb!]
	\centering
        		\includegraphics[width=1\textwidth]{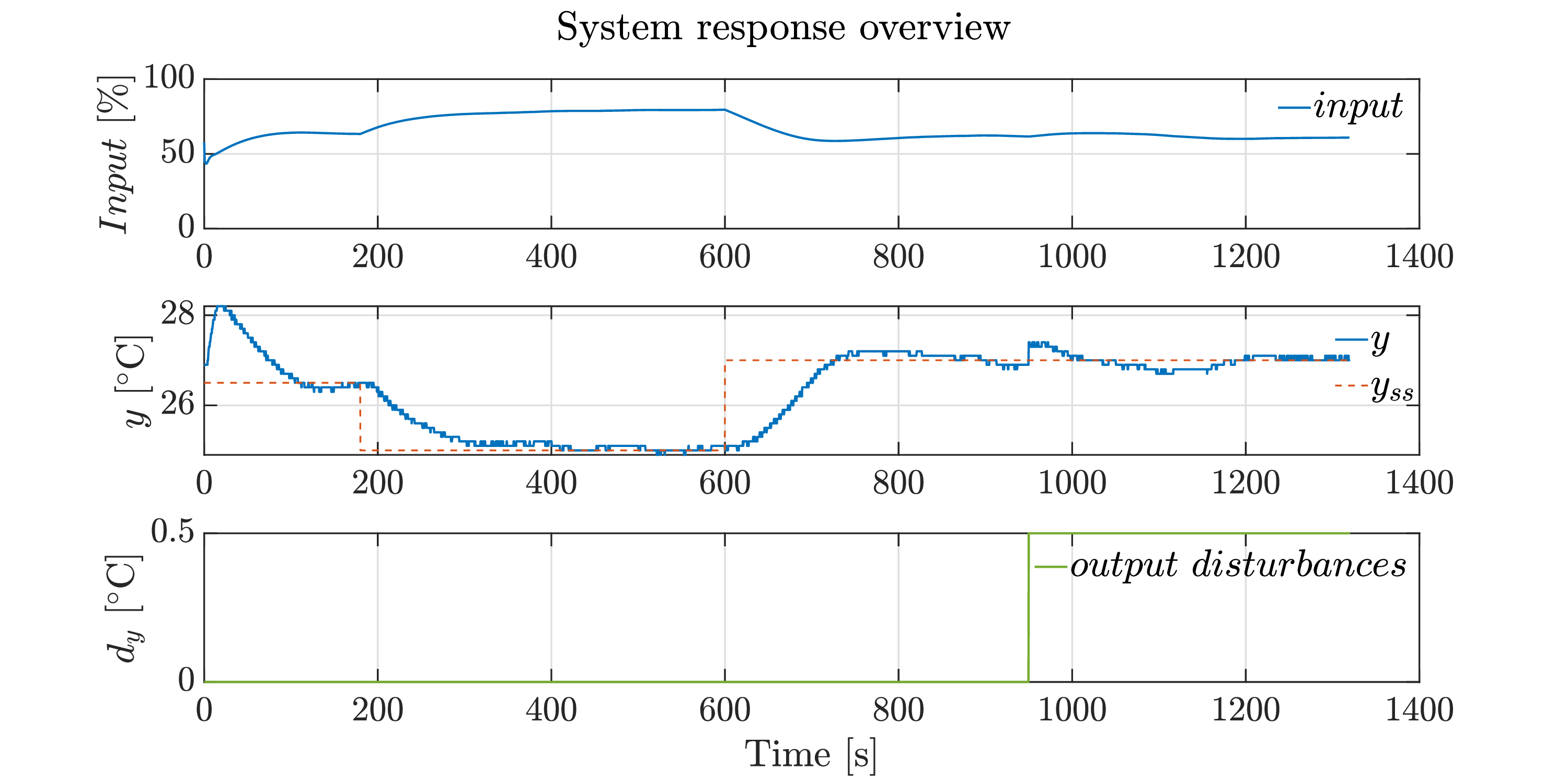}
\caption{Subfigures arranged vertically (from top to bottom) showing: the input signal, the system output, and the output disturbance.}\label{fig:first_exp}
\end{figure*}

\begin{figure*}[htb!]
	\centering
        		\includegraphics[width=1\textwidth]{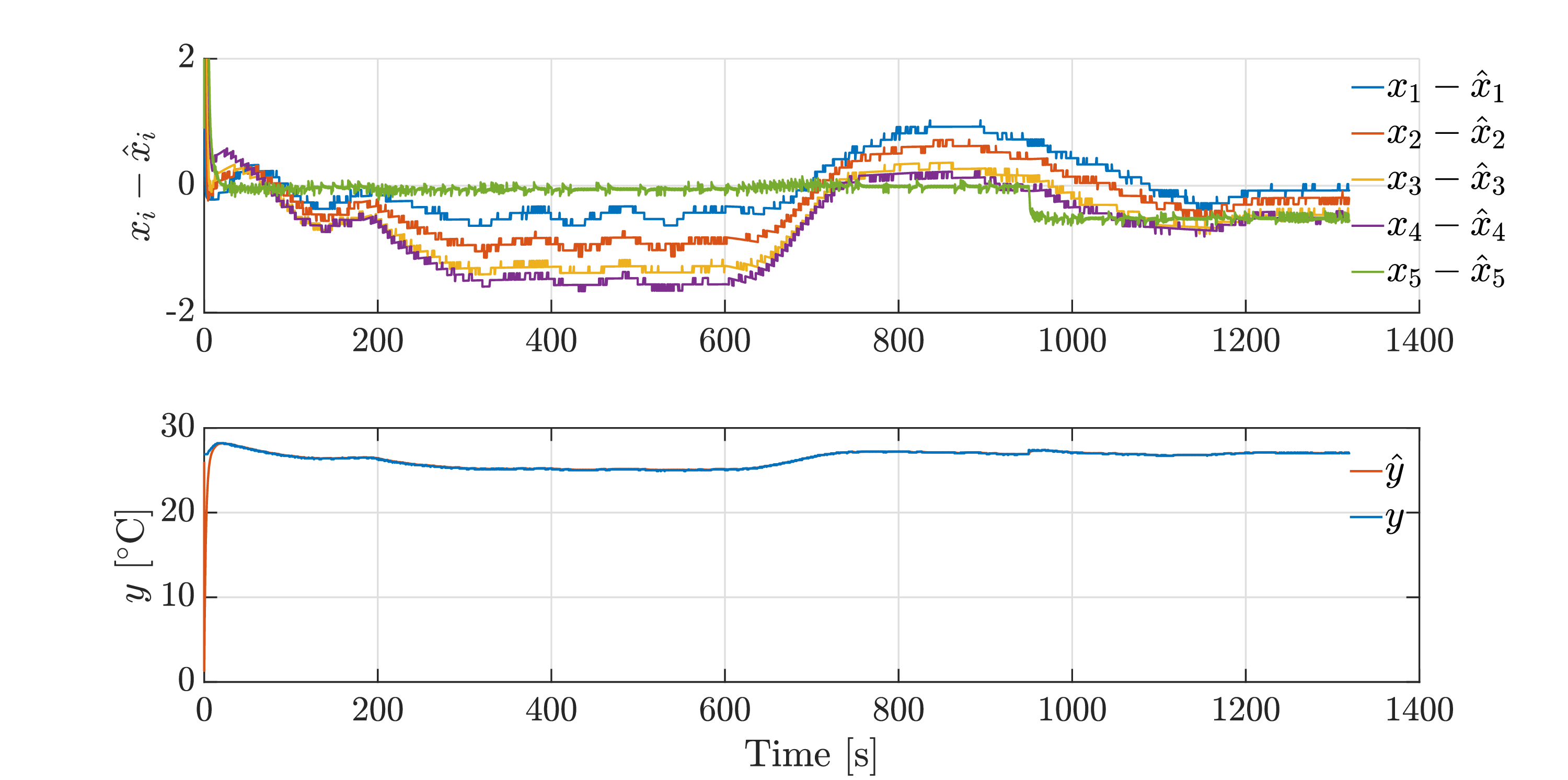}
    \caption{The first subfigure shows the observation error; the second subfigure displays both the system output and the estimated output.}
\label{fig:first_exp_part2}
\end{figure*}

These results underscore a fundamental advantage of the proposed control strategy: the ability to accurately reconstruct the full temperature profile $\hat{x}$ along the cold fluid channel, despite the availability of only a limited number of physical sensors. This is made possible by the observer integrated in the  control scheme, which benefits from global convergence guarantees to reliably estimate unmeasured internal states.
The maximum observed error is approximately one degree Celsius, which can largely be attributed to the uncertainty in the alignment between the physical sensors and the discretization grid used in the model. Given this source of discrepancy, such a small deviation represents a strong validation of the observer’s accuracy.
This capability holds significant practical value in industrial contexts, where sensor deployment may be constrained by cost, accessibility, or physical space. Being able to infer the full system state from sparse measurements not only improves monitoring and diagnostics but also enables more precise and robust control. 

\subsection{Second experiment}

In the second experiment, we compare our proposed controller with a first-order PI controller. \textcolor{black}{To ensure a fair comparison, the PI gains $k_p$ and $k_i$ were selected to optimize the performances of the system when the command input is around 80\% of its maximum value. This choice is motivated by the fact that, in this nonlinear framework, no systematic tuning method is available that simultaneously guarantees stability and performance over the whole operating range. A model-based tuning based on a linearization around a single equilibrium point would not be appropriate here, since our goal is to allow the system to operate across different equilibria, for instance in scenarios where the reference temperature changes frequently.}


As in the first experiment, we analyze their performance under varying reference conditions. Specifically, the reference temperature is changed from 26.5°C to 26°C after 240 seconds, then to 28°C at 550 seconds, and finally to 24.4°C after 900 seconds.
From figure \eqref{fig:second_experiment}, it is evident that our controller never reaches the saturation zone, instead stabilizing at a maximum of 80\% of its saturation value. In contrast, the PI controller becomes fully saturated after 950 seconds and remains in this state.
Regarding reference tracking, our controller exhibits a slight overshoot of 0.5°C around 650 seconds, which is not observed in the PI controller. However, as we move further from the PI’s linear operating region, its performance deteriorates significantly. In fact, even after 300 seconds (5 minutes) from a reference change, the PI controller still fails to reach the desired value, effectively behaving as if it were operating in open-loop mode.
These results confirm the improved performance of the proposed controller based on two key aspects. First, it is supported by a theoretical framework that ensures global convergence and closed-loop stability, unlike the PI controller, which lacks such guarantees. Second, the control input corresponds to the valve opening that regulates the water flow rate. By avoiding saturation, the proposed controller allows for a lower average flow rate. In this experiment, the valve remains below 80\% opening, while the PI controller reaches full saturation. This difference corresponds to an approximate 20\% reduction in water usage.
By combining formal stability guarantees with efficient resource utilization, the proposed control strategy proves especially well-suited for industrial applications where both high performance and minimized consumption are essential.

\begin{figure*}[htb!]
    \centering
    		\includegraphics[width=1\textwidth]{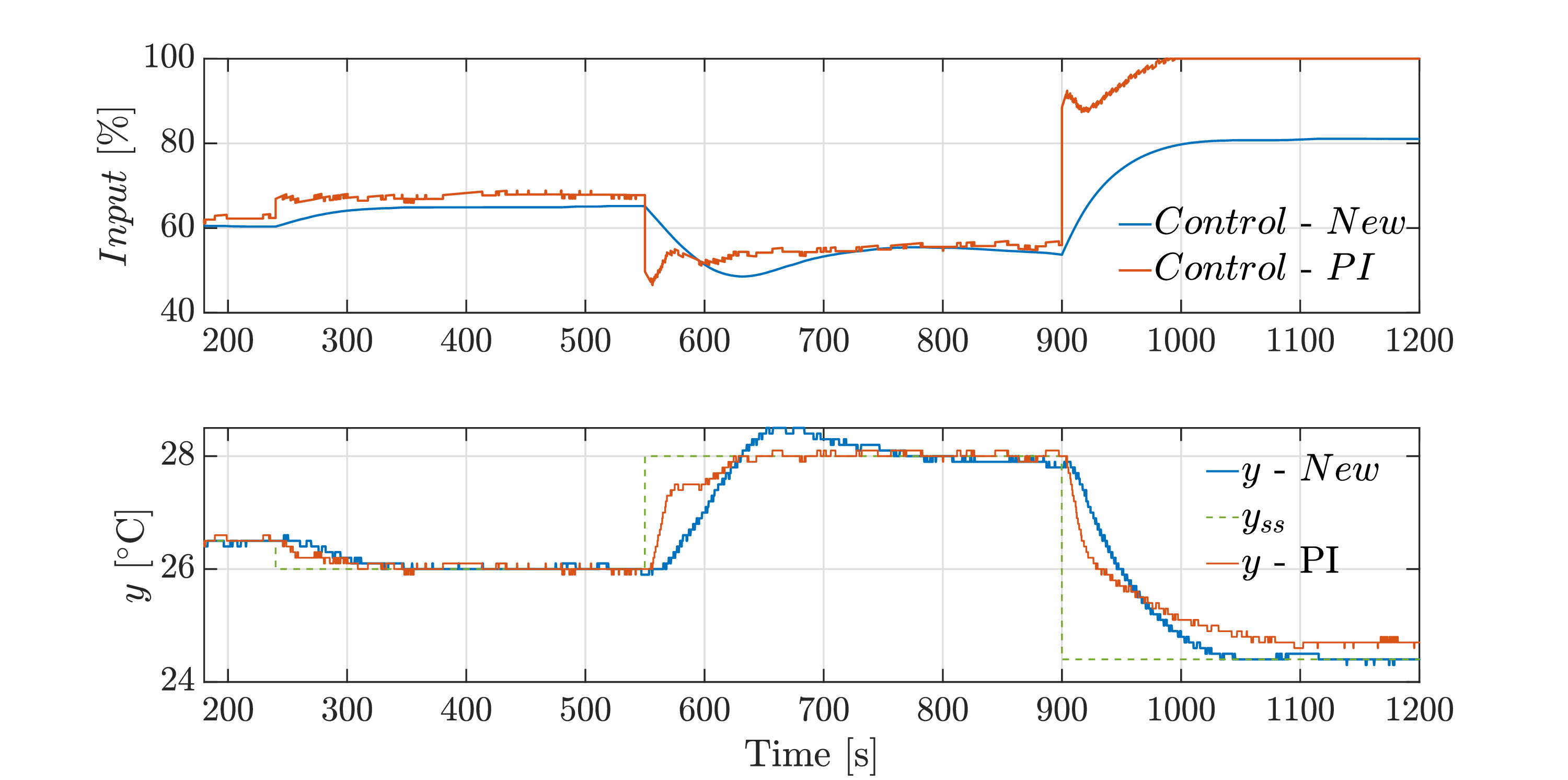}
    
    \caption{}
    \label{fig:second_experiment}
\end{figure*}

\section{Conclusion and perspectives}\label{sec:conclusions}

We developed innovative theoretical frameworks focusing on output constant reference tracking for single-input single-output bilinear systems in the presence of (possibly asymmetric) input saturation.
{\color{black}
To address the regulation problem for bilinear systems with intrinsic input saturation, two 
integral-action based
output-feedback control strategies were developed. 
The first strategy combines an integral action and a  state feedback design with a state observer through a separation-principle framework, explicitly tailored to account for the bilinear structure of the system and to ensure practical tunability. 
The second strategy relies on a pure output-feedback law directly combined with integral action, offering a simpler implementation at the cost of more restrictive theoretical assumptions.
The proposed methodology is formulated under structural conditions commonly satisfied by several physical processes, including heat exchangers. 
In both strategies, the control design explicitly incorporates the intrinsic saturation affecting the system dynamics, thereby reflecting the physical limitations imposed by real actuators.
The observer-based strategy was experimentally validated on a real heat exchanger and benchmarked against a standard proportional–integral (PI) controller widely adopted in industrial practice. 
Experimental results demonstrate the effectiveness and robustness of the proposed control architecture, highlighting its advantages over conventional linear control approaches.
}
A natural follow-up of this work is the exploration of robust output regulation for finite-dimensional nonlinear systems, particularly in the bilinear setting \cite{bin2022robustness}. Addressing this problem would make it possible to handle more complex reference signals or disturbances, including time-varying or periodic ones, expanding beyond the constant-reference scenarios tackled through integral action in this study. One promising direction involves leveraging infinite-dimensional internal model structures, such as repetitive control schemes \cite{astolfi2021repetitive}, for the robust tracking of periodic signals.

Concerning potential applications, a particularly interesting challenge is the distributed control of a network of heat exchangers, as found in district heating systems \cite{dobos2009dynamic} or large-scale industrial processes. This could initially be tackled by formulating it as a synchronization problem for bilinear systems \cite{li2009consensus,arcak2007passivity}, a research area where many questions remain open. Eventually, a distributed integral control strategy \cite{andreasson2014distributed}, or more complex internal model-based solutions, could be adopted to regulate the temperature of each exchanger to a common reference profile.

\bibliographystyle{ieeetr}
\bibliography{biblio}

\end{document}